\documentclass[pra,11pt,onecolumn,notitlepage]{revtex4-2}

\usepackage{amsmath}
\usepackage{mathtools}
\usepackage{multirow}
\usepackage{bbm, dsfont}
\usepackage[colorlinks, citecolor=blue]{hyperref}
\usepackage[utf8]{inputenc}
\usepackage{graphicx}
\usepackage{caption, subcaption}
\newcommand{\RN}[1]{%
  \textup{\uppercase\expandafter{\romannumeral#1}}%
}
\begin{document}
	
\title{Beyond Qubits : An Extensive Noise Analysis for Qutrit Quantum Teleportation}
	\author{Arun Sebastian}
	\email{aruns04_2023@cmscollege.ac.in}
	\affiliation{%
			Department of Physics, CMS College, Kottayam, Kerala 686001, India
	}%
	\author{Afnan. N. Mansar}
        \email{manzarafnan3011@gmail.com}
	\affiliation{%
        Department of Physics, TKM College Of Arts \& Science, 
        Karicode, Peroor, Kollam, Kerala 691005, India
	}%
	\author{N. C. Randeep}
        \email{randeepvarnam@gmail.com}
	\affiliation{%
	Department of Physics, Government Arts and Science College, Meenchanda, Calicut, Kerala 673018, India
	}%
	\date{\today}
\begin{abstract}
  The four quantum noises Bit Flip, Phase Flip, Depolarization, and Amplitude Damping as well as any potential combinations of them are examined in this paper's investigation of quantum teleportation using qutrit states. Among the above mentioned noises, we observed phase flip has highest fidelity. Compared to uncorrelated Amplitude Damping, we find that correlated Amplitude Damping performs two times better. Finally, we agreed that, for better fidelity, it is preferable to provide the same noise in channel state if noise is unavoidable. 
\end{abstract}
\maketitle
\section{Introduction}
The discipline of quantum information science has undergone notable advancements recently\cite{MOL2017, TOT2014,GIS2002,MA2016}, advancing our comprehension of the quantum environment and paving the way for the creation of ground-breaking technology. Quantum tasks such as quantum teleportation \cite{BEN1993}, super dense coding \cite{BEN1992}, quantum key distribution \cite{BEN2014}, quantum error correction \cite{DEV2013} and remote state preparation \cite{BEN2005} lead to this revolution. Underlying principle behind these astonishing developments is quantum entanglement \cite{CHU2009,HOR2009}, that correlate two or more particles no matter how far apart they are. Out of these developments, quantum teleportation is the most famous one that attracted scientific researcher most. It is a non-classical technique for sending quantum information over great distances and was originally presented in 1993 by Bennett et al. \cite{BEN1993}. Its experimental implementation was done by S. Popescu, A. Zeilinger and their groups in 1997 \cite{BOU1997}.  Initially quantum teleportation was proposed for qubits systems, later it has been developed for higher dimensional systems (qudits) \cite{MUT2000,KAS2000}.  Particularly, teleportation using qutrit (3 dimensional qudit) systems have been studied in detail \cite{HUA2020}. Experimentally, different degree of freedoms such as spatial modes\cite{KRE2021,GOM2021}, frequency modes\cite{OLI2012,BER2013,JIN2016}, orbital angular momentum\cite{DAD2011,FIC2012,ERH2018} and time bins\cite{MAR2017,IKU2017} have been used to develop entangled qudits. Perfect teleportation is possible only when there exist a maximum entanglement between sender and receiver, also efficiency of teleportation is maximum in this case. Efficiency decreases with decrease in degree of entanglement between sender and receiver. The efficiency of teleportation can be measured by a quantity called fidelity $(F)$, which is $1$ for maximally entangled state and is $0$ for not entangled state \cite{OH2002}. Theoretically in all teleportation scheme we are using maximally entangled state and we obtain the maximum fidelity $F=1$. But in real situations, due to interactions with the environment, the entanglement between particles gradually weaken or vanish altogether over time. Since entanglement is a delicate and fragile characteristic that is quickly disrupted by a variety of factors, such as noise, decoherence, or interactions with outside particles which may leads to entanglement sudden death \cite{YU2006,YU2009,ANN2008,ALI2010}. So in real situation we cannot expect maximum fidelity of teleportation $(F=1)$ and it decreases with noises. Many attempts have been made to study various noise effects in different teleportation schemes \cite{FOR2015,BAD2000,YEO2008,BHA2008,KNO2014,FOR2016,FON2019}. In which Fortes and Rigolin in $2015$ discussed a situation in which more noise leads to more efficiency in the case of qubit teleportation \cite{FOR2015}.
In this paper we analyse noise effects on qutrit teleportation scheme by applying various combinations of noises such as bit flip, amplitude damping, phase damping and depolarization channel etc on input state and entangled channel. We observe that noise decreases the fidelity of teleportation compared to noise less case. On the other hand in certain situations such as adding more noises and introducing correlation effect increases the fidelity of teleportation.

This paper assembled as follows, Section \ref{sec:Review of Qutrit teleportation} of this work discusses the qutrit teleportation protocol through density matrix formalism. Modelling of noise and the system noise interaction and the efficiency of noisy teleportation protocol are covered in Section \ref{sec:Noise modelling and efficiency in the noisy protocol}. In part \ref{sec:Constant noise acting only in Alice input state and Bob's channel state effected by all other four noise.}, we investigate fidelity when noise only affects Alice's input state, and in section \ref{sec:Constant noise acting in Alice input qutrit and channel qutrit  and Bob's channel qutrit effected by all other noises.}, we investigate fidelity when noise affects both Alice state. The effect of noise in both channel states and noise abstraction is included in Section \ref{sec:Constant noise acting in channel state} and Section \ref{sec:Correlated Amplitude Damping(CAD).} shows the correlated amplitude damping and normal amplitude in channel state gives the same effect on qutrit quantum teleportation.
\maketitle \section{Review of Qutrit teleportation} \label{sec:Review of Qutrit teleportation}
Qubits are two level quantum system which can be considered as the superposition of two orthonormal states  $|0 \rangle$ and $|1 \rangle$ and are basic building block of quantum information. An extended version of qubits to higher dimension \cite{HUA2020} is also possible. In three dimension which is qutrits, in four which is ququads and in general $N$  dimension which is qudits. Qudits offers more secure and powerful quantum computations. Generally a qutrit state is of the form,$$$$
\begin{equation}
    |\psi\rangle = a|0\rangle + b|1\rangle + c|2\rangle
\label{eq:qutritinput}
\end{equation}
where $|0\rangle $, $|1\rangle $ and $|2\rangle $ are elements of computational basis set for qutrits and in matrix notation which can be expressed as 
\begin{equation}
    | 0\rangle = \begin{pmatrix}
    1\\ 0\\ 0 \end{pmatrix}, 
    | 1\rangle  = \begin{pmatrix} 0\\ 1\\ 0 \end{pmatrix}, | 2\rangle =\begin{pmatrix} 0\\ 0\\ 1 \end{pmatrix}
\end{equation}
also $a,b$ and $c$ are the complex coefficients which satisfies the normalisation condition,
\begin{equation}
 |a|^2 + |b|^2 + |c|^2 = 1
\end{equation}
Now, using density matrix framework, we analyse standard quantum teleportation \cite{MOL2017}  for a single unknown Qutrit state via two Qutrit entangled states. Consider Alice's input single Qutrit state is as mentioned in Eq.~(\ref{eq:qutritinput}), we can write it in terms of density matrix as follows, 
\begin{equation} 
\rho_{in} =  |\psi_{in}\rangle \langle\psi_{in}| =  \begin{pmatrix}
a^2 & ab^* & ac^*\\
ba^* & b^2 & bc^*\\
ca^* & cb^* & c^2 
\end{pmatrix},
\end{equation}
where the subscript “in” means “input” and “*” denotes complex conjugation.
Let us suppose that Alice and Bob share a two qutrit maximally entangled channel in which the first qutrit is with Alice and the second one is with Bob,
\begin{equation} 
    |\psi _{ch}\rangle = \frac{1}{\sqrt 3} ( |00 \rangle + |11 \rangle + |22 \rangle ),
\end{equation}
and the corresponding density matrix can be written as  
\begin{equation} 
\rho_{ch} = |\psi_{ch}\rangle \langle \psi_{ch}|,  
\end{equation}
which is a 9 $\times$ 9 matrix with the following non zero elements as follows
\begin{equation*}
    \rho_{ch_{1,1}} = \rho_{ch_{1,5}} = \rho_{ch_{1,9}} = \rho_{ch_{5,1}} = \rho_{ch_{5,5}} =\rho_{ch_{5,9}}=\rho_{ch_{9,1}} = \rho_{ch_{9,5}} =\rho_{ch_{9,9}} =\frac{1}{\sqrt{3}}
\end{equation*}
Then the total density matrix of the system is the tensor product of input density matrix $\rho_{in}$ and channel density matrix $\rho_{ch}$,
\begin{equation} 
    \rho = \rho_{in} \otimes \rho_{ch} 
    \label{totaldenmatrix}
\end{equation}
To proceed further, Alice wants to make a projective measurement on her qutrit  in which she projects her qutrit into following $9$ maximally entangled states, which forms the basis for two qutrit system:
\begin{align}
\begin{split}
|\phi_1\rangle &= \frac{1}{\sqrt3}(|00\rangle + |11\rangle + |22\rangle )\\
|\phi_2\rangle &= \frac{1}{\sqrt3}(|00\rangle + e^{i \frac{2\pi}{3}} |11\rangle + e^{i \frac{4\pi}{3}} |22\rangle )\\
|\phi_3\rangle &= \frac{1}{\sqrt3}(|00\rangle + e^{i \frac{4\pi}{3}} |11\rangle + e^{i \frac{2\pi}{3}}|22\rangle )\\
|\phi_4\rangle &= \frac{1}{\sqrt3}(|01\rangle + |12\rangle + |20\rangle )\\
|\phi_5\rangle &= \frac{1}{\sqrt3}(|01\rangle + e^{i \frac{2\pi}{3}}|12\rangle + e^{i \frac{4\pi}{3}} |20\rangle )\\
|\phi_6\rangle &= \frac{1}{\sqrt3}(|01\rangle + e^{i \frac{4\pi}{3}} |12\rangle + e^{i \frac{2\pi}{3}}|20\rangle )\\
|\phi_7\rangle &= \frac{1}{\sqrt3}(|02\rangle + |10\rangle + |21\rangle )\\
|\phi_8\rangle &= \frac{1}{\sqrt3}(|02\rangle + e^{i \frac{2\pi}{3}} |10\rangle + e^{i \frac{4\pi}{3}} |21\rangle )\\
|\phi_9\rangle &= \frac{1}{\sqrt3}(|02\rangle + e^{i \frac{4\pi}{3}} |10\rangle + e^{i \frac{2\pi}{3}} |21\rangle ).
\end{split}
\label{totalentangledstate}
\end{align}
The projection operator associated with each basis can be written as 
\begin{equation} 
P_j  = |\phi_j\rangle\langle \phi_j|
\label{Projective}
\end{equation}
Alice perform a joint measurement using the projective measurement operator in Eq.(\ref{Projective}) on the total density matrix given by  Eq.(\ref{totaldenmatrix}) using anyone of basis in Eq.(\ref{totalentangledstate}) and obtain the density matrix.
\begin{equation}
\tilde{\rho_{j}} = \frac{P_j \rho P_j^\dagger}{Tr[P_j\rho]},
\end{equation}
where $Tr[P_j\rho]$ gives the probability of the occurrence of the particular  $\Tilde{\rho_{j}}$.
After Alice measurement she inform the resultant $|\phi _j\rangle $  to Bob via a classical channel and he obtain corresponding state,
\begin{equation}
    \rho^\sim_{B_j} = Tr_{12}[\Tilde{\rho_{j}}] = \frac{Tr_{12}[P_j \rho P_j^\dagger]}{Tr[P_j\rho]},
\end{equation}
where $Tr_{12}$ is the partial trace operation of Alice's two qutrits and resulting to the density matrix of Bob's state.
Finally Bob does a unitary operation $U_{j}$ on his state $\rho^\sim_{B_j}$ and he can reproduce input state send by Alice. Which is of the form,
\begin{equation}\label{eq:12}
\rho_{B_j} = U_j\rho^{\sim}_{B_j}U_j^{\dagger}.
\end{equation}
In fact the operation $U_j$ convert the final Bob's state $\Tilde{\rho_{B_j}}$ to initial input state $\rho_{in}$ with a unit fidelity. Eq.(\ref{eq:12}) gives us unit fidelity because  teleportation protocol we consider is an ideal one that is we consider a closed quantum system. Now we consider real system in which the noise acts on different states and we examine the deviation of protocol efficiency from unity.

\section{Noise modelling and efficiency in the noisy protocol} \label{sec:Noise modelling and efficiency in the noisy protocol}

Mainly there are four types of standard noises that is possible to act on qutrit namely bit flip, phase flip, amplitude damping and depolarizing noise. The state of the system $\rho$ evolve in the presence of noise and become $\epsilon(\rho)$, in the operator sum representation which can be written as 

\begin{equation}
\epsilon (\rho)=\sum \limits_{j}  K_{j} \rho K_{j}^{\dagger}
\end{equation}
where $K_{j}$s are Kraus operators for the given noise. These Kraus operators satisfies the normalization condition $\sum\limits_{j} K_{j} K_{j}^{\dagger}=\mathds{1}$ and are different for different types of noises. Next we can look at the forms these Kraus operators for above mentioned noises. 

\subsection{Bit Flip Noise}

In quantum information processing a Bit Flip noise refers to noise that act on each quantum state and flip into another state. In our problem it act on discrete qutrit state and flip into another qutrit state with a finite probability $p$. Mathematically it can be thought as following
\begin{equation}
\begin{split}
|0\rangle \xrightarrow{} |1\rangle, |1\rangle \xrightarrow{} |2\rangle, |2\rangle \xrightarrow{} |0\rangle \\
|0\rangle \xrightarrow{} |2\rangle, |1\rangle \xrightarrow{} |0\rangle, |2\rangle \xrightarrow{} |1\rangle.
\end{split}
\end{equation}
It's Kraus operators have the following form, 
\begin{equation} 
K_{0} = \sqrt{1-p} \mathds{1}, \hspace{0.5cm} K_1 = \sqrt{\frac{p}{2}}\begin{pmatrix}
0 & 0 & 1\\
1 & 0 & 0\\
0 & 1 & 0 
\end{pmatrix}, \hspace{0.5cm} K_2 = \sqrt{\frac{p}{2}}\begin{pmatrix}
0 & 1 & 0\\
0 & 0 & 1\\
1 & 0 & 0 
\end{pmatrix}.
\end{equation}

\subsection{Phase Flip Noise}
Phase flip noise effects the phase of the qutrit, which changes qutrit state with probability $p$ as follows,   
$|1\rangle \xrightarrow{} -|1\rangle, |2\rangle \xrightarrow{} -|2\rangle$. Then the corresponding Kraus operator can be written as

\begin{equation} \label{eq2} 
K_{0} = \sqrt{1-p} \mathds{1}, \hspace{0.5cm} K_1 = \sqrt{\frac{p}{2}}\begin{pmatrix}
1 & 0 & 0\\
0 & -1 & 0\\
0 & 0 & 1 
\end{pmatrix}, \hspace{0.5cm} K_2 = \sqrt{\frac{p}{2}}\begin{pmatrix}
1 & 0 & 0\\
0 & 1 & 0\\
0 & 0 & -1 
\end{pmatrix}.
\end{equation}
\subsection{Depolarizing Noise}
Depolarizing noise causes the quantum state  of a qubit to become a completely mixed state with a certain probability. The Depolarizing noise Kraus operators for qutrits have the following form,
\vspace{10pt}
\begin{equation} 
K_{0} = \sqrt{1-p} \mathds{1}, \hspace{0.5cm} K_1 = \sqrt{\frac{p}{8}} Y, \hspace{0.5cm} K_2 = \sqrt{\frac{p}{8}}Z, \hspace{0.5cm} K_3 = \sqrt{\frac{p}{8}}Y^2, \hspace{0.5cm} K_4 = \sqrt{\frac{p}{8}}YZ
\end{equation}
\begin{equation}
K_5 = \sqrt{\frac{p}{8}}Y^2Z, \hspace{0.5cm} K_6 = \sqrt{\frac{p}{8}}YZ^2, \hspace{0.5cm} K_7 = \sqrt{\frac{p}{8}}Y^2Z^2, \hspace{0.5cm} K_8 = \sqrt{\frac{p}{8}}Z^2
\end{equation}

where  $Y= \begin{pmatrix}
0 & 1 & 0\\
0 & 0 & 1\\
1 & 0 & 0 
\end{pmatrix}$, $Z = \begin{pmatrix}
1 & 0 & 0\\
0 & \omega & 0\\
0 & 0 & \omega^2 
\end{pmatrix}$ and $\omega = e^{i \frac{2\pi}{3}}$
\subsection{Amplitude Damping Noise}
The Amplitude Damping noise is an important noise which is responsible for the energy dissipation of quantum system for the upper level\cite{yeo2003time,grassl2018quantum}. Our consideration is a three level system in which it have three 
different type of configuration namely $
\vee, \Lambda, 	\Xi$ (ladder system) configurations\cite{schirmer2004constraints}. Energy dissipation for each configuration govern by different set of Kraus operator's. We only focus on $\vee$ system. The Kraus operators for the $\vee$ system can be written as

\begin{equation}  
K_{0} = \begin{pmatrix}
1 & 0 & 0\\
0 & \sqrt{1-p_1} & 0\\
0 & 0 & \sqrt{1-p_2}
\end{pmatrix}, \hspace{0.5cm} K_1 = \begin{pmatrix}
0 & \sqrt{p_1} & 0\\
0 & 0 & 0\\
0 & 0 & 0 
\end{pmatrix}, \hspace{0.5cm} K_2 = \begin{pmatrix}
0 & 0 & \sqrt{p_2}\\
0 & 0 & 0\\
0 & 0 & 0 
\end{pmatrix}
\end{equation} 
\vspace{10pt}

Where $p_1 = 1 - e^{-\gamma_1t}$ and $p_2 = 1 - e^{-\gamma_2t}$, are strength of amplitude damping and $\gamma_1$, $\gamma_2$ are life time of upper state $|1\rangle$ and $|2\rangle$, respectively.\vspace{20pt}

Now its the time to understand how teleportation is effected by the noises. The total density matrix at the starting of teleportation is given by Eq.(\ref{totaldenmatrix}), and different possible noises act in different ways on this state. Action of most general noise such as acting noise on input state and Alice and Bob parts of the channel, can be expressed as 

\begin{equation}
    \varrho=\sum_{i=0}^{n_I}E_i(p_i)\left[\sum_{j=0}^{n_A}F_j(p_A)\left[\sum_{k=0}^{n_B}G_k(p_B) \rho G_k^\dagger(p_B)\right]{F_j^\dagger(p_A)}\right]E_i^\dagger(p_I)
    \end{equation} 
    \begin{equation}   =\sum_{i=0}^{n_I}\sum_{j=0}^{n_A}\sum_{k=0}^{n_B}K_{ijk}(p_I ,p_A ,p_B)\rho K_{ijk}^\dagger(p_I ,p_A ,p_B)
   \end{equation}
where $E_i(p_I)$ ,$F_j(p_A)$ and $G_k(p_B)$ are the Kraus operators of noise acting on the input state , Alice's channel state  and Bob's channel  state with probabilities $ p_I,p_A $ and $p_B$.\newpage Also $K_{ijk}(p_I,p_A,p_B) =E_i(p_I)\otimes F_j(p_A)\otimes G_k(p_B)$ and  $E_i(p_I) = E_i(p_I)\otimes \mathds{1} \otimes \mathds{1} $  ,$F_j(p_A) = \mathds{1} \otimes F_j(p_A) \otimes \mathds{1}$ , 
 $G_k(p_B)=\mathds{1}\otimes \mathds{1}\otimes G_k(p_B)$.
   
   \vspace{10pt}
 The efficiency of the teleportation protocol is given by the quantity referred as Fidelity. For pure state the fidelity is given by
 \begin{equation}
     F_j = Tr[\rho_{in}\varrho_{B_j}] = \langle\psi|\varrho_{B_j}|\psi\rangle
 \end{equation}
 where $\varrho_{B_j}$ is given by eqn (12) in which $\rho$ is replaced by $\varrho$ in the noisy state.
 The average fidelity \cite{li2019enhance}can be obtained by
 \begin{equation}
     F_{av} = \frac{1}{4\pi}\int_{0}^{\pi}d\theta\int_{0}^{2\pi}d\phi F(\theta,\phi)sin \theta
 \end{equation}

We will now examine how the noises described above behave in various states and affect teleportation effectiveness. We demonstrate the impact of combining various noise channels with different states.  

\section{Constant noise acting only in Alice input state and Bob's channel state effected by all other four noise.}\label{sec:Constant noise acting only in Alice input state and Bob's channel state effected by all other four noise.}
In the last section we have discussed about four different noises. These different noises may act many ways in the input state and channel. In this section we are going discuss how a constant noise acting on the input state is  effected by various noises acting on the Bob's part of the channel.
\subsection{Noise acting only in Alice input state}

First we can discuss most simplest one, noise acting on Alice's input state only. That is channel state is protected from all other noise. For each four types of noise the average fidelity can be written as,

\begin{equation}
\langle F_{BF, non, non}\rangle = 1 - \frac{4p_I}{5}
\end{equation}
\begin{equation}
\langle F_{PF, non, non}\rangle = 1 - \frac{8p_I}{15}
\end{equation}
\begin{equation}
\langle F_{DP, non, non}\rangle = 1 - \frac{3p_I}{5}
\end{equation}
\begin{equation}
\langle F_{AD, non, non}\rangle = - \frac{2p_I}{5} + \frac{4\sqrt{1-p_I}}{15} + \frac{11}{15}
\end{equation}

where each term on the left hand side of above four expressions $\langle F_{Noise, non, non}\rangle$ denote noises namely BF(Bit Flip), PF(Phase Flip), DP(Depolarizing), AD(Amplitude Damping) act on Alice input state with probability p and noise free in the Alice part and Bob part of the channel.

\begin{figure}[ht!]
    \centering
    \includegraphics[width=0.5\textwidth]{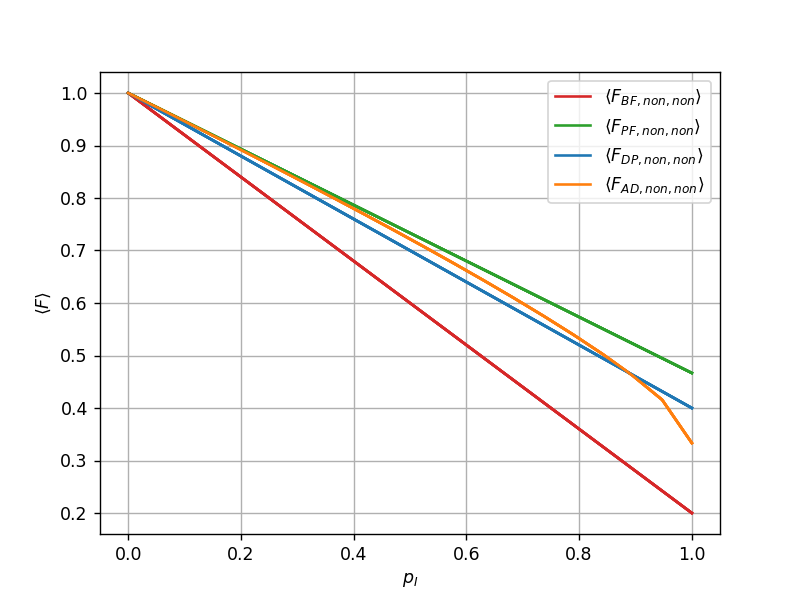}
    \caption{Fidelity v/s probability plot. It shows the Fidelity of teleportation when four different noise acting only in Alice input state}
    \label{Figure1}
\end{figure}

In the FIG.~(\ref{Figure1}), we demonstrate the variation of average fidelity with respect to various noises with probability $p$. In all the four cases, rate of decrease  for average fidelity is maximum for Bit flip case and minimum for phase flip case. Other two lies in between.

Next we consider what happens to fidelity when constant noise acting only in Alice input state and different noises acting on Bob's part of channel state. In this case we assume that Alice part of the channel state is protected from all noises.

\subsection{Constant Bit Flip acting on Alice input state and Different noise acting on Bob's state}

Now we consider what happens when a constant Bit Flip act on Alice input state with probability $p_I$ and different noise act on Bob's channel state with probability $p_B$. The fidelity in each case can be written as,

\begin{equation}
\langle F_{BF, non, BF}\rangle = \frac{6p_Ip_B}{5} - \frac{4p_I}{5} - \frac{4p_B}{5} + 1
\end{equation}
\begin{equation}
\langle F_{BF, non, PF}\rangle = \frac{8p_Ip_B}{15} - \frac{4p_I}{5} - \frac{8p_B}{15} + 1
\end{equation}
\begin{equation}
\langle F_{BF, non, DP}\rangle = \frac{9p_Ip_B}{10} - \frac{4p_I}{5} - \frac{3p_B}{5} + 1
\end{equation}
\begin{equation}
\langle F_{BF, non, AD}\rangle = \frac{8p_Ip_B}{15} - \frac{8p_I}{15} - \frac{2p_B}{5} + \frac{4\sqrt{1-p_B}}{15} - \frac{4p_I\sqrt{1-p_B}}{15} + \frac{11}{15}
\end{equation}

\begin{figure}[ht!]
\centering
    \begin{subfigure}[b]{0.3\textwidth}
        \includegraphics[width=\textwidth]{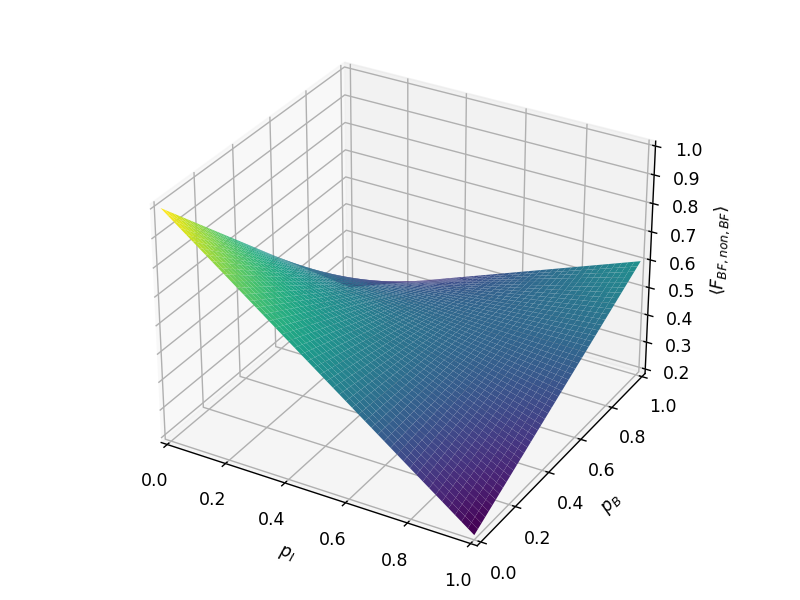}
        \caption{$\langle F_{BF,non,BF} \rangle$ as a function of $p_I$ and $p_B$ }
        \label{fig 2(a)}  
    \end{subfigure}
    \hfill
    \begin{subfigure}[b]{0.3\textwidth}
        \includegraphics[width=\textwidth]{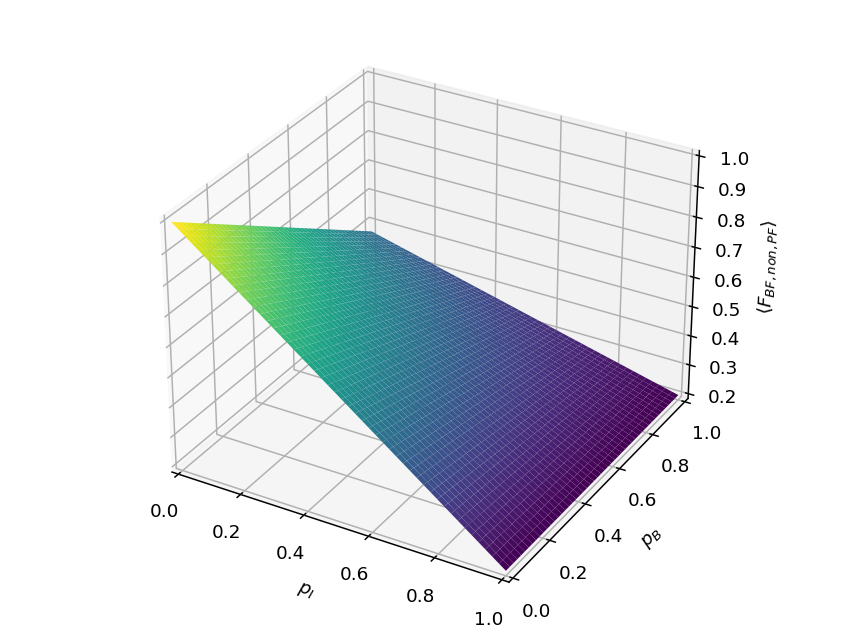}
        \caption{$\langle F_{BF,non,PF} \rangle$ as a function of $p_I$ and $p_B$}
        \label{fig 2(b)}  
    \end{subfigure}
    \hfill
    \begin{subfigure}[b]{0.3\textwidth}
        \includegraphics[width=\textwidth]{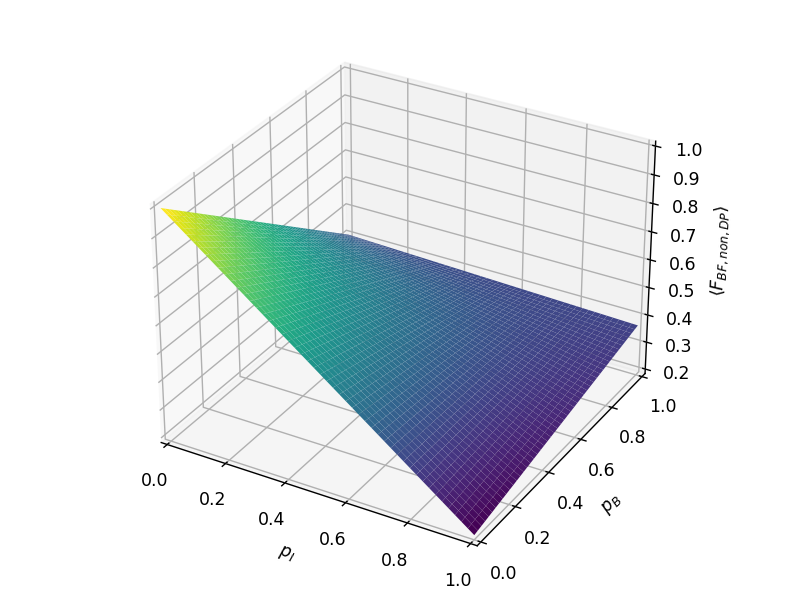}
        \caption{$\langle F_{BF,non,DP} \rangle$ as a function of $p_I$ and $p_B$}
        \label{fig 2(c)}
    \end{subfigure}
    \hfill
    \begin{subfigure}[b]{0.3\textwidth}
        \includegraphics[width=\textwidth]{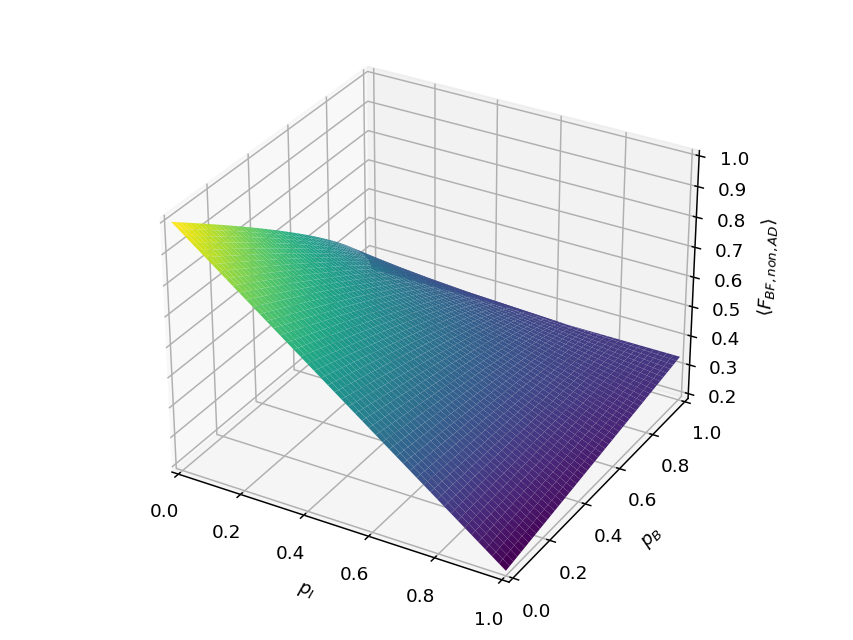}
        \caption{$\langle F_{BF,non,AD} \rangle$ as a function of $p_I$ and $p_B$}
        \label{fig 2(d)}
    \end{subfigure}
    \hfill
    \begin{subfigure}[b]{0.3\textwidth}
        \includegraphics[width=\textwidth]{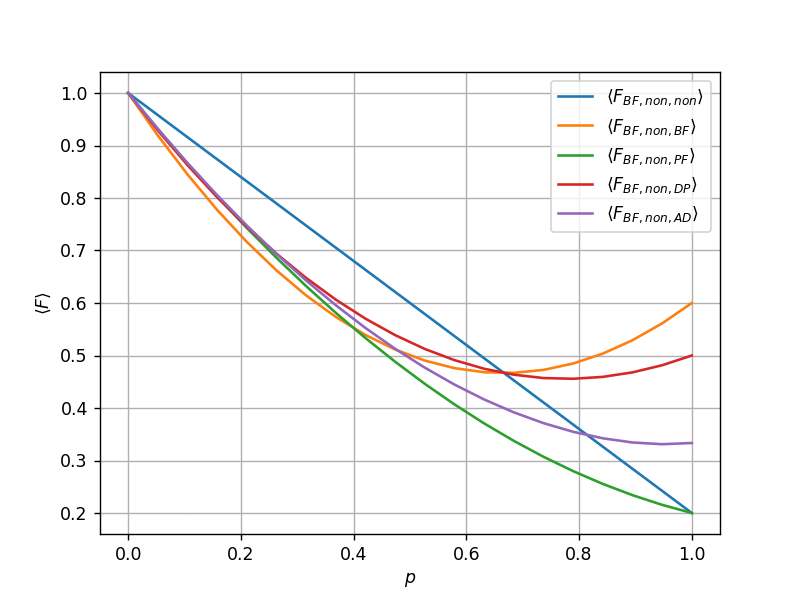}
        \caption{a comparison of FIG 2(a), FIG 2(b), FIG 2(c), FIG 2(d) as $p_I = p_B = p$ }
        \label{fig 2(e)}
    \end{subfigure}
\caption{Plot of constant Bit Flip acting on Alice input state and different noise acting on Bob's part of the channel state}
\label{Figure2}
\end{figure}

FIG.~(\ref{Figure2}), shows the variation of average fidelity with probabilities $p_{I}$ and $p_{B}$. When the input bit flip noise strength is maximum ($p_I = 1 $) and noise in the Bob's part of the channel is absent ($p_{B}=0$), then the fidelity is $\frac{1}{5}$. While introducing $p_{B}$ in this case, increases the average fidelity with $p_{B}$ in the case of bit flip, depolarising and amplitude damping noises, but with phase flip noise fidelity remains constant. That is we observe that noise plus noise increases the fidelity of teleportation at least in the  three situations discussed above. By taking $p_{I}=p_{B}=p$ and plotted average fidelity with $p$ as shown in FIG.~(\ref{fig 2(e)}) and its values when $p=1$ are related by $\langle F_{BF,non,PF}\rangle = \langle F_{BF,non,non}\rangle < \langle F_{BF,non,AD}\rangle < \langle F_{BF,non,DP}\rangle < \langle F_{BF,non,BF}\rangle$.

\subsection{Constant Phase Flip acting on Alice input state and Different noise acting on Bob's state}
Next consider Phase Flip acts on Alice input state with probability $p_{I}$ and different noise act on Bob’s channel state with probability $p_B$ . The fidelity in each case can be written as
\begin{equation}
\langle F_{PF, non, BF}\rangle = \frac{8p_Ip_B}{15} - \frac{8p_I}{15} -\frac{4p_B}{5} + 1
\end{equation}
\begin{equation}
\langle F_{PF, non, PF}\rangle = \frac{32p_Ip_B}{45} - \frac{8p_I}{15} -\frac{8p_B}{15} + 1
\end{equation}
\begin{equation}
\langle F_{PF, non, DP}\rangle = \frac{2p_Ip_B}{5} - \frac{8p_I}{15} - \frac{3p_B}{5} + 1
\end{equation}
\begin{equation}
\langle F_{PF, non, AD}\rangle = \frac{8p_Ip_B}{45} - \frac{8p_I}{45} - \frac{2p_B}{5} + \frac{4\sqrt{1-p_B}}{15} - \frac{16p_I\sqrt{1-p_B}}{45} + \frac{11}{15}
\end{equation}
\begin{figure}[ht!]
\centering
    \begin{subfigure}[b]{0.3\textwidth}
        \includegraphics[width=\textwidth]{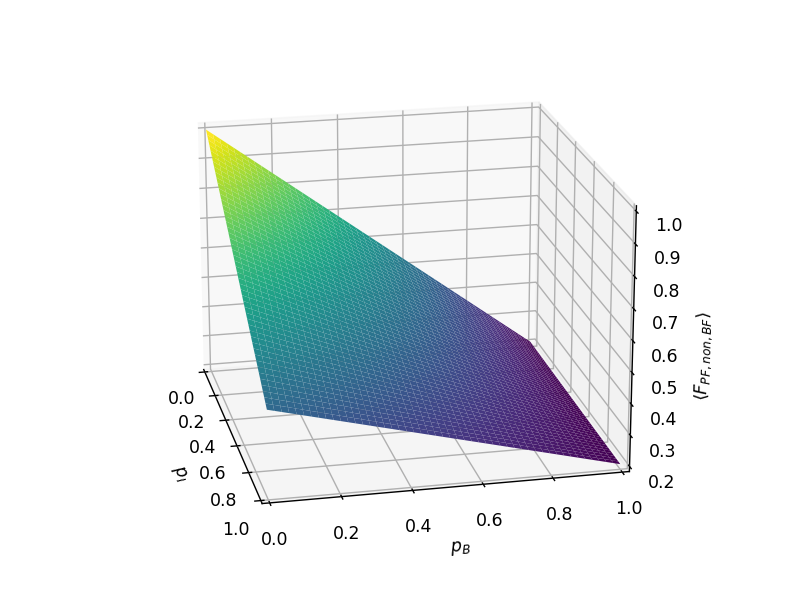}
        \caption{$\langle F_{PF,non,BF} \rangle$ as a function of $p_I$ and $p_B$}
        \label{fig 3(a)}
    \end{subfigure}
    \hfill
    \begin{subfigure}[b]{0.3\textwidth}
        \includegraphics[width=\textwidth]{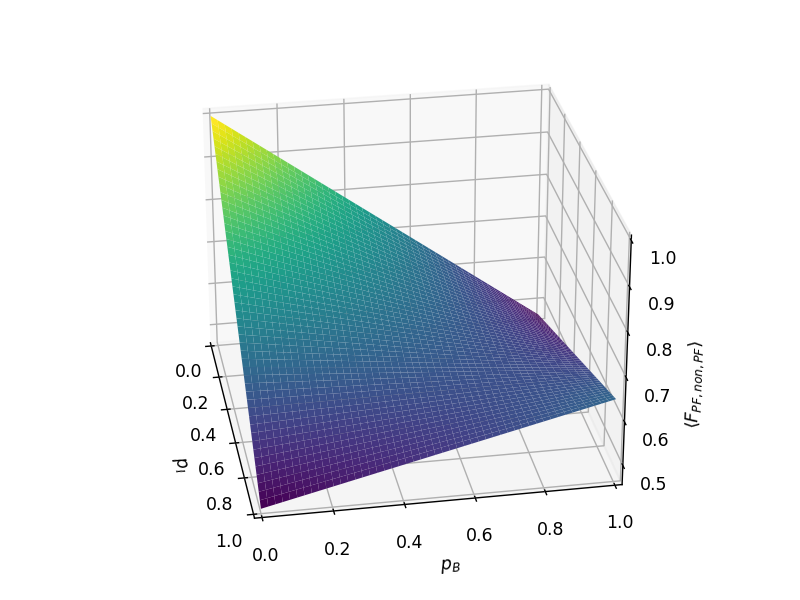}
        \caption{$\langle F_{PF,non,PF} \rangle$ as a function of $p_I$ and $p_B$ }
        \label{fig 3(b)}
    \end{subfigure}
    \hfill
    \begin{subfigure}[b]{0.3\textwidth}
        \includegraphics[width=\textwidth]{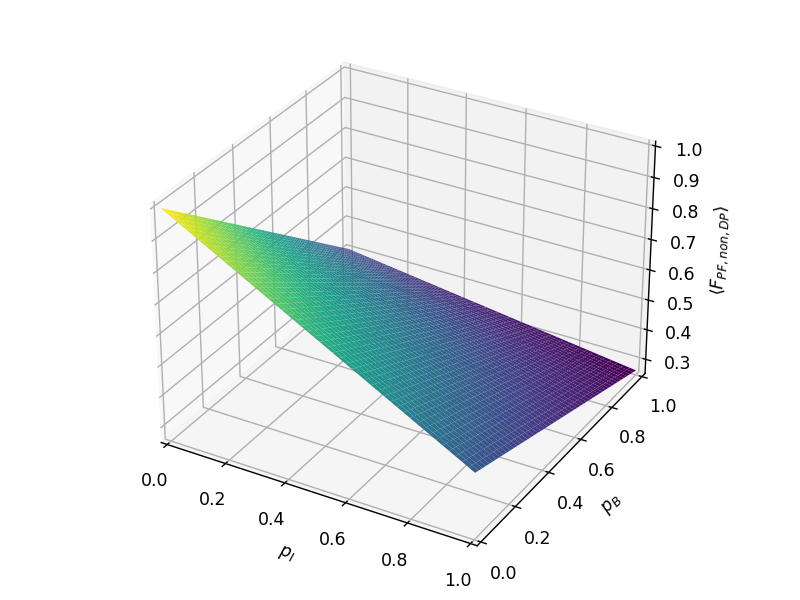}
        \caption{$\langle F_{PF,non,DP} \rangle$ as a function of $p_I$ and $p_B$}
        \label{fig 3(c)}
    \end{subfigure}
    \hfill
    \begin{subfigure}[b]{0.3\textwidth}
        \includegraphics[width=\textwidth]{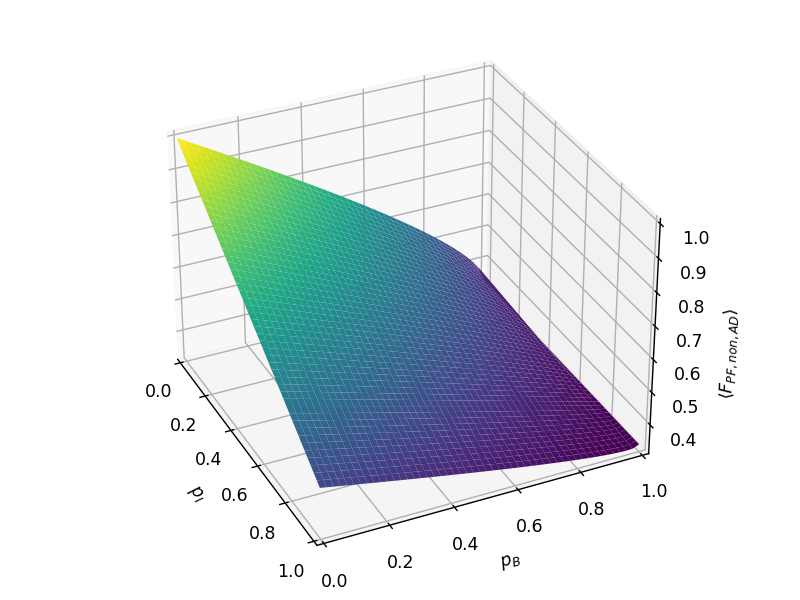}
        \caption{$\langle F_{PF,non,AD} \rangle$ as a function of $p_I$ and $p_B$}
        \label{fig 3(d)}
    \end{subfigure}
    \hfill
    \begin{subfigure}[b]{0.3\textwidth}
        \includegraphics[width=\textwidth]{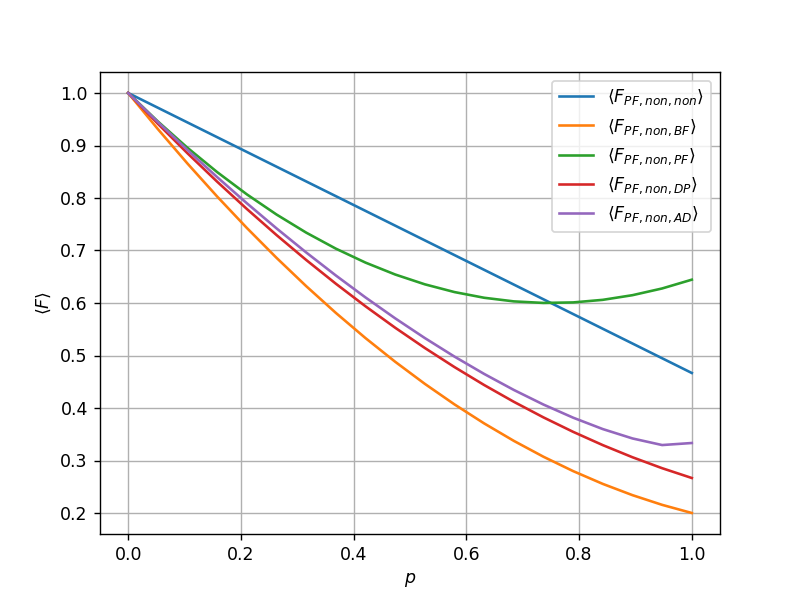}
        \caption {a comparison of FIG 3(a), FIG 3(b), FIG 3(c), FIG 3(d) as $p_I = p_B = p$}
        \label{fig 3(e)}
    \end{subfigure}
\caption{Constant Phase Flip acting on Alice input state and different noise acting on Bob's state}
\label{Figure3}
\end{figure}

In FIG.~(\ref{Figure3}), as in the last section we have studied variation of average fidelity with $p_{I}$ and $p_{B}$. Here $p_{I}$ indicate phase flip and $p_{B}$ s indicate all the four noises. From FIG.~(\ref{fig 3(e)}), we consider the variation of average fidelity with $p_{B}$  when $p_{I}=1$, we observe that average fidelity decreases with $p_{B}$ except for phase flip noise. In the phase flip noise, average fidelity increases with $p_{B}$. In a single FIG.~(\ref{fig 3(e)}) , we showed the variation of all four noises with $p_{I}=p_{B}=p$. The average fidelity when $p=1$ for different noises are related by   $\langle F_{PF,non,BF}\rangle < \langle F_{PF,non,DP}\rangle < \langle F_{PF,non,AD}\rangle < \langle F_{PF,non,non}\rangle < \langle F_{PF,non,PF}\rangle$.
\subsection{Constant Depolarizing Noise acting on Alice input state and Different noise acting on Bob's state}
In this case depolarizing noise acts on Alice input state with probability $p_{I}$ and different noises act on Bob's part of the channel with probability $p_{B}$. Then the average fidelity in different cases are 
\begin{equation}
\langle F_{DP, non, BF}\rangle = \frac{9p_Ip_B}{10} - \frac{3p_I}{5} - \frac{4p_B}{5} + 1
\end{equation}
\begin{equation}
\langle F_{DP, non, PF}\rangle = \frac{2p_Ip_B}{5} - \frac{3p_I}{5} - \frac{8p_B}{15} + 1
\end{equation}
\begin{equation}
\langle F_{DP, non, DP}\rangle = \frac{27p_Ip_B}{40} - \frac{3p_I}{5} - \frac{3p_B}{5} + 1
\end{equation}
\begin{equation}
\begin{split}
\langle F_{DP, non, AD}\rangle = \frac{p_I\sqrt{1-p_B}}{15 } -\frac{2p_I}{5}+ \frac{2p_B(p_I-1)}{5} +\frac{4(1-p_I)\sqrt{1-p_B}}{15} + \frac{11}{15}
\end{split}
\end{equation}

\begin{figure}[ht!]
\centering
    \begin{subfigure}[b]{0.3\textwidth}
        \includegraphics[width=\textwidth]{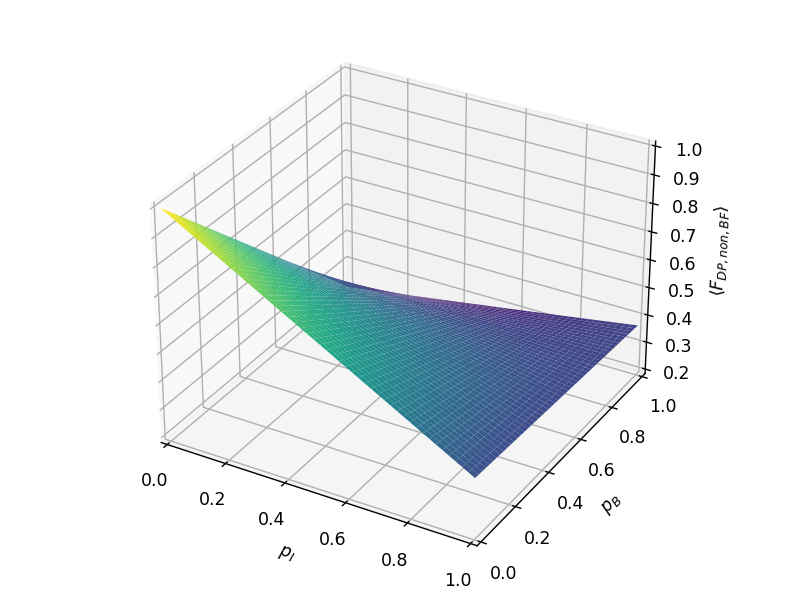}
        \caption{$\langle F_{DP,non,BF} \rangle$ as a function of $p_I$ and $p_B$}
        \label{fig 4(a)}
    \end{subfigure}
    \hfill
    \begin{subfigure}[b]{0.3\textwidth}
        \includegraphics[width=\textwidth]{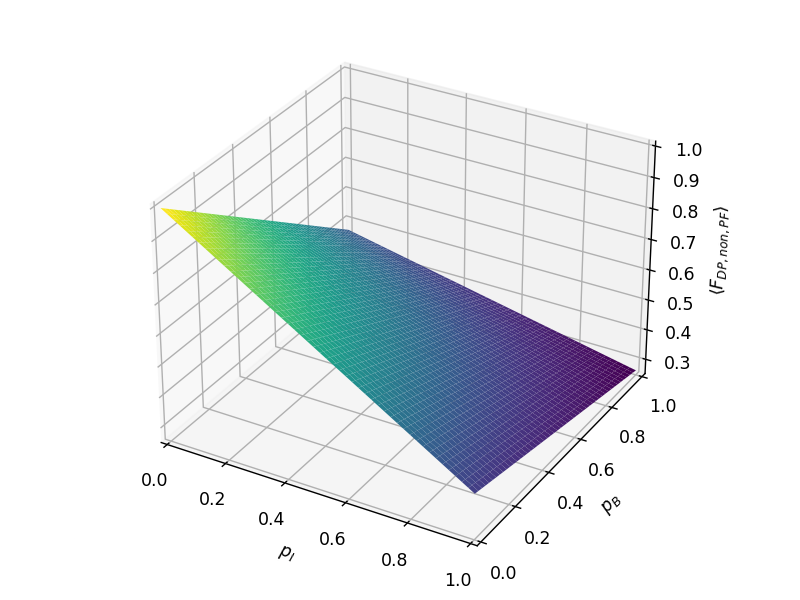}
        \caption{$\langle F_{DP,non,PF} \rangle$ as a function of $p_I$ and $p_B$}
        \label{fig 4(b)}
    \end{subfigure}
    \hfill
    \begin{subfigure}[b]{0.3\textwidth}
        \includegraphics[width=\textwidth]{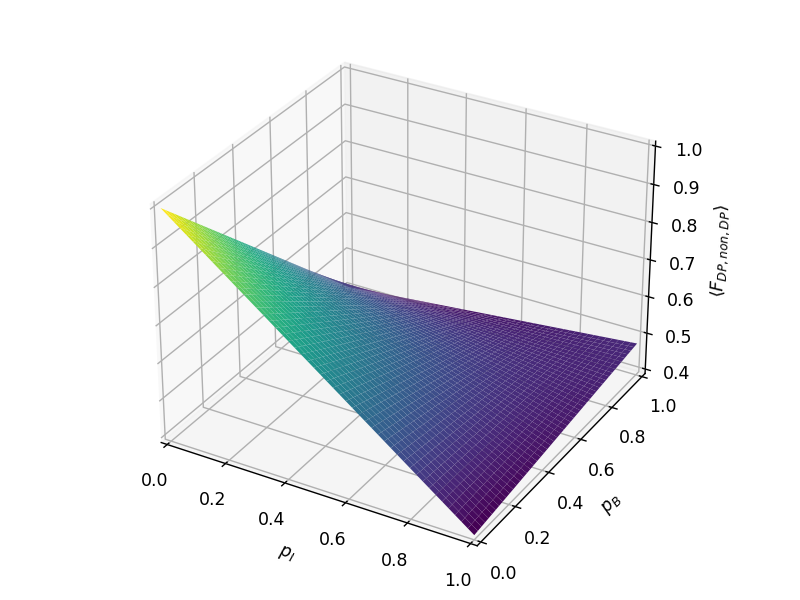}
        \caption{$\langle F_{DP,non,DP} \rangle$ as a function of $p_I$ and $p_B$}
        \label{fig 4(c)}
    \end{subfigure}
    \hfill
    \begin{subfigure}[b]{0.3\textwidth}
        \includegraphics[width=\textwidth]{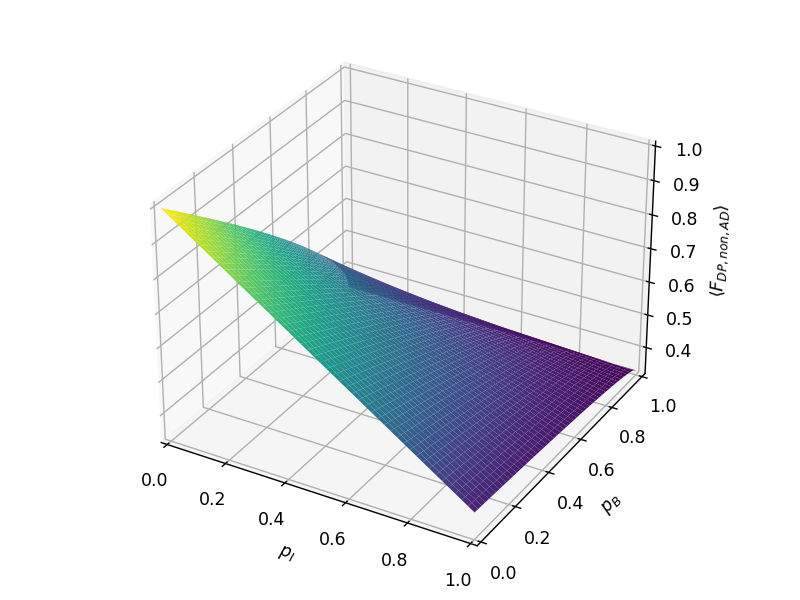}
        \caption{$\langle F_{DP,non,AD} \rangle$ as a function of $p_I$ and $p_B$}
        \label{fig 4(d)}
    \end{subfigure}
    \hfill
    \begin{subfigure}[b]{0.3\textwidth}
        \includegraphics[width=\textwidth]{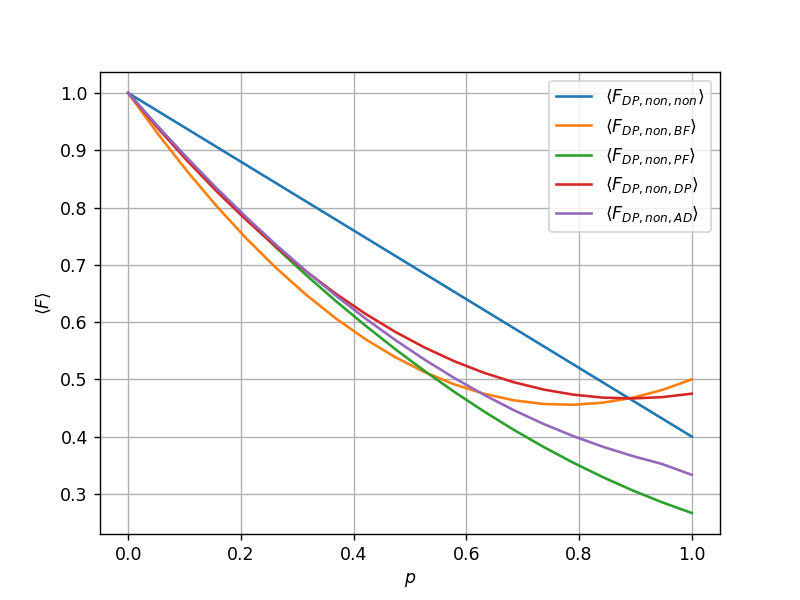}
        \caption{a comparison of FIG 4(a), FIG 4(b), FIG 4(c), FIG 4(d) as $p_I = p_B = p$}
        \label{fig 4(e)}
    \end{subfigure}
\caption{Constant Depolarizing Noise acting on Alice input state and Different noise acting on Bob's state}
\label{Figure4}
\end{figure}
In FIG.~(\ref{Figure4}), we choose input noise with probability $p_{I}$ and noise in the Bob's part of the channel with probability $p_{B}$. While considering the variation of average fidelity  with $p_{B}$ for $p_{I}=1$, we observe that fidelity decreases in the case of phase flip and amplitude damping in the Bob's part of the channel but increases in the case of bit flip and depolarising noises in the same. Also FIG.~(\ref{fig 4(e)}) shows the variation of all four combination of noises with $p_{B}=p_{I}=p$. The average fidelity when $p=1$ for different noises can be sorted as   $\langle F_{DP,non,PF}\rangle < \langle F_{DP,non,AD}\rangle < \langle F_{DP,non,non}\rangle < \langle F_{DP,non,DP}\rangle < \langle F_{DP,non,BF}\rangle$.

\subsection{Constant Amplitude Damping on Alice input state and different noise acting on Bob's state}

Finally consider Amplitude Damping($\vee$ system)  acting on Alice input qutrit channel and other four noise acting on the Bob's channel state. The average fidelity in different cases have the following form. \newpage
\begin{equation}
\langle F_{AD, non, BF}\rangle = \frac{8p_Ip_B}{15} - \frac{2p_I}{5} - \frac{4p_B\sqrt{1-p_I}}{15} - \frac{8p_B}{15} + \frac{4\sqrt{1-p_I}}{15} + \frac{11}{15}
\end{equation}
\begin{equation}
\langle F_{AD, non, PF}\rangle = \frac{8p_Ip_B}{45} - \frac{2p_I}{5} - \frac{16p_B\sqrt{1-p_I}}{45} - \frac{8p_B}{45} + \frac{4\sqrt{1-p_I}}{15} + \frac{11}{15}
\end{equation}
\begin{equation}
\langle F_{AD, non, DP}\rangle = \frac{19p_Ip_B}{30} - \frac{2p_I}{5} - \frac{7p_B^2}{30} - \frac{p_B\sqrt{1-p_I}}{5} - \frac{2p_B}{5}  + \frac{4\sqrt{1-p_I}}{15} + \frac{11}{15}
\end{equation}
\begin{equation}
\begin{split}
\langle F_{AD, non, AD}\rangle = \frac{14p_Ip_B}{45} - \frac{4p_I\sqrt{1-p_B}}{45} - \frac{14p_I}{45} - \frac{4p_B\sqrt{1-p_I}}{45} - \frac{14p_B}{45} + \\ \frac{8\sqrt{1-p_I}\sqrt{1-p_B}}{45} + \frac{4\sqrt{1-p_I}}{45} + \frac{4\sqrt{1-p_B}}{45} + \frac{29}{45}
\end{split}
\end{equation}

\begin{figure}[ht!]
\centering
    \begin{subfigure}[b]{0.3\textwidth}
        \includegraphics[width=\textwidth]{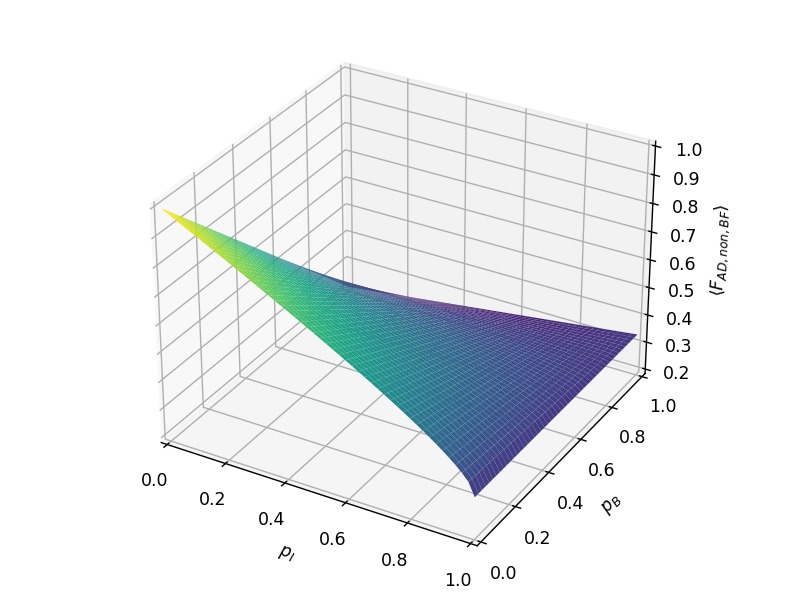}
        \caption{$\langle F_{AD,non,BF} \rangle$ as a function of $p_I$ and $p_B$}
        \label{fig 5(a)}
    \end{subfigure}
    \hfill
    \begin{subfigure}[b]{0.3\textwidth}
        \includegraphics[width=\textwidth]{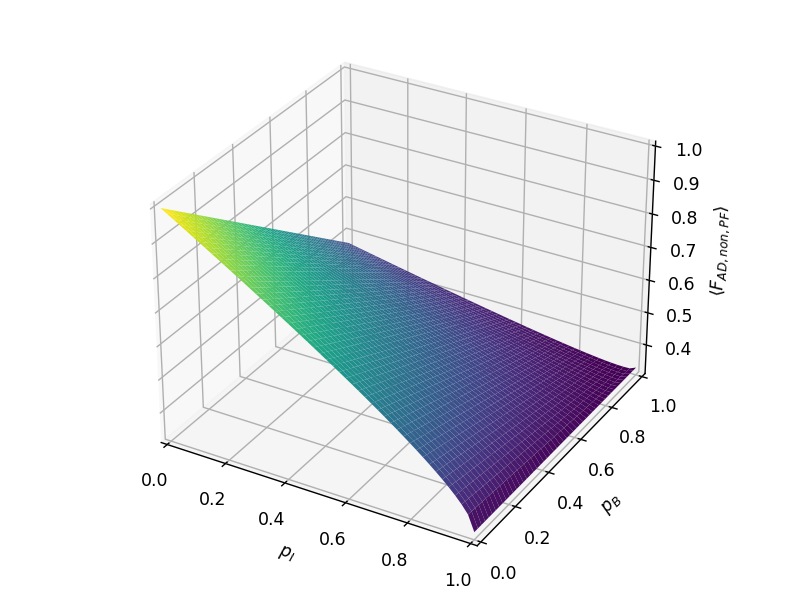}
        \caption{$\langle F_{AD,non,PF} \rangle$ as a function of $p_I$ and $p_B$}
        \label{fig 5(b)}
    \end{subfigure}
    \hfill
    \begin{subfigure}[b]{0.3\textwidth}
        \includegraphics[width=\textwidth]{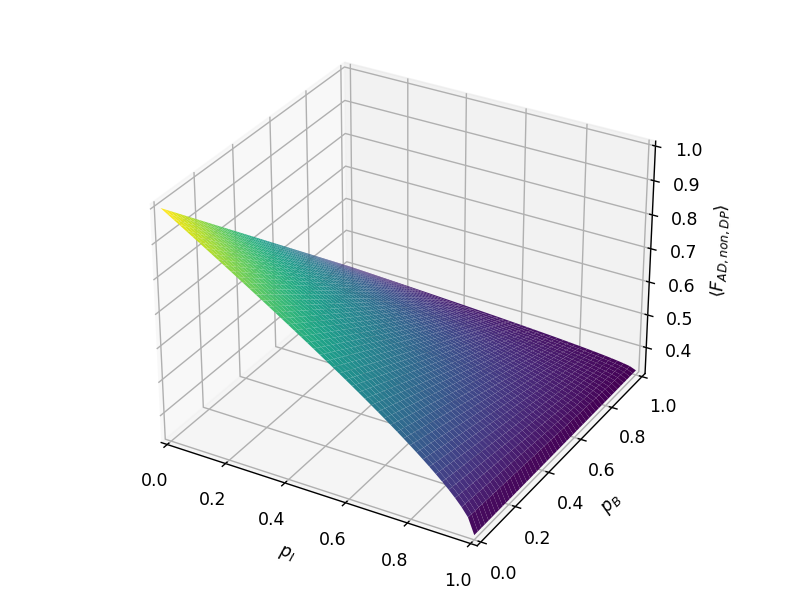}
        \caption{$\langle F_{AD,non,DP} \rangle$ as a function of $p_I$ and $p_B$}
        \label{fig 5(c)}
    \end{subfigure}
    \hfill
    \begin{subfigure}[b]{0.3\textwidth}
        \includegraphics[width=\textwidth]{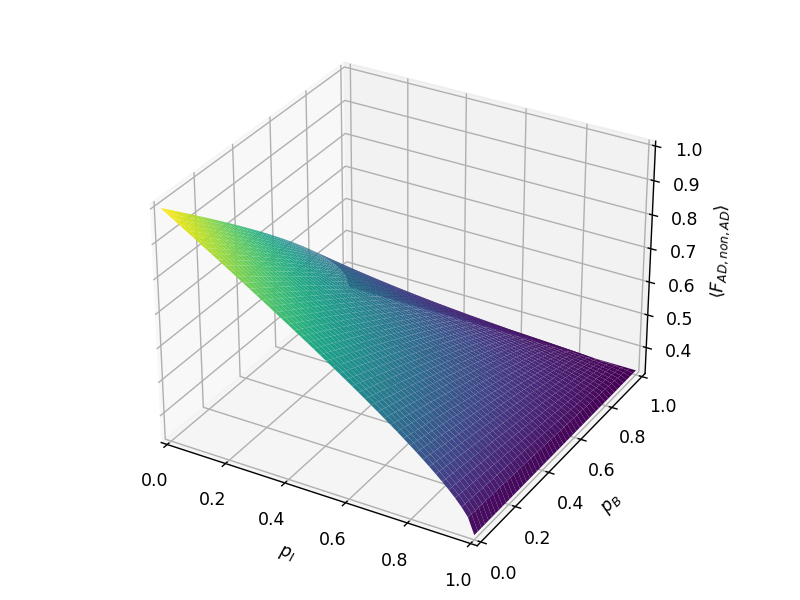}
        \caption{$\langle F_{AD,non,AD} \rangle$ as a function of $p_I$ and $p_B$}
        \label{fig 5(d)}
    \end{subfigure}
    \hfill
    \begin{subfigure}[b]{0.3\textwidth}
        \includegraphics[width=\textwidth]{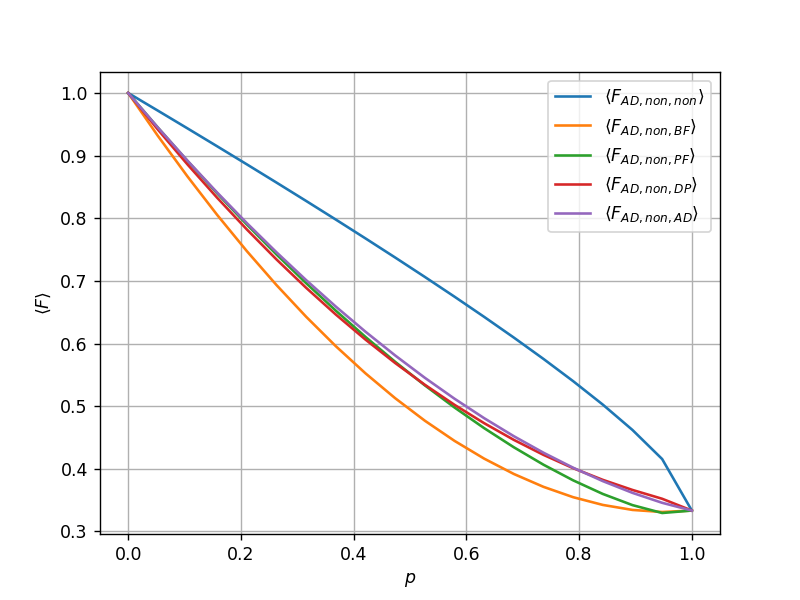}
        \caption{a comparison of FIG 5(a), FIG 5(b), FIG 5(c), FIG 5(d) as $p_I = p_B = p$}
        \label{fig 5(e)}
    \end{subfigure}
\caption{Constant Amplitude Damping on Alice input state and different noise acting on Bob's state}
\label{Figure5}
\end{figure}

FIG.(\ref{Figure5}) shows the variation of average fidelity with $p_{I}$ and $p_{B}$. In which average fidelity is independent of $p_{B}$  when $p_{I}=1$. Also in FIG.(\ref{fig 5(e)}) illustrates an extremely intriguing result: regardless of the noise that is operating on Bob’s channel state, when constant amplitude damping is act on Alice’s input state, we always obtain a constant minimum fidelity when both the probability is at its highest.
\section{Constant noise acting in Alice input qutrit and channel qutrit  and Bob's channel qutrit effected by all other noises.}\label{sec:Constant noise acting in Alice input qutrit and channel qutrit  and Bob's channel qutrit effected by all other noises.}
Now let us consider the cases when noise acting on both Alice state, and Bob's channel state is effected by all other noise.  
\subsection{Noise acting on Alice input state and channel state and   Bob's state is isolated from noise}
First, we study the case when Bob's channel state is isolated from all noises and constant noise acts in Alice's both states. The average fidelity of teleportation in such case can be written as follows
\begin{equation}
\langle F_{BF, BF , non}\rangle = \frac{6p^2}{5} - \frac{8p}{5}  + 1 
\end{equation}
\begin{equation}
\langle F_{PF, PF , non}\rangle = \frac{32p^2}{45} - \frac{16p}{15}  + 1 
\end{equation}
\begin{equation}
\langle F_{DP, DP , non}\rangle = \frac{27p^2}{40} - \frac{6p}{5}  + 1 
\end{equation}
\begin{equation}
\langle F_{AD, AD , non}\rangle = \frac{14p^2}{45} - \frac{8p\sqrt{1-p}}{45}  -\frac{4p}{5} +\frac{8\sqrt{1-p}}{45} + \frac{37}{45} 
\end{equation}
\begin{figure}[ht!]
    \centering
    \includegraphics[width = 0.5\linewidth]{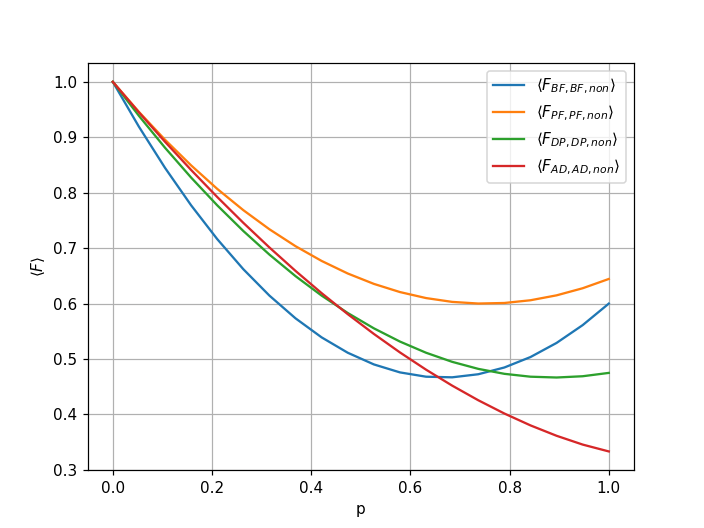}
    \caption{shows average fidelity when constant noise act in input state and Alice's channel state ($p_I = p_A = p$) and Bob's channel excluded from noise. }
    \label{figure6}
\end{figure}
In FIG.~(\ref{figure6}), we demonstrate the change of average fidelity of the case considered above. The decrease of average fidelity is maximum for amplitude damping and minimum for phase flip. It is worth noticing that average fidelity is greater in this case than in the case when noise acts only in Alice input state. This can be verified by comparing FIG.(\ref{Figure1}) and FIG.(\ref{figure6})

\subsection{Constant Bit Flip acting on Alice input state and channel state and Different noise acting on Bob's state}
Let's see what happens when a constant Bit Flip acts on Alice input state and Alice channel state with probabilities $p_I$ and $p_A$  respectively and different types of noise act on Bob's channel state with probability $p_B$. The average fidelity in each case can be written as,
\begin{equation}
\begin{split}
\langle F_{BF, BF, BF}\rangle = -\frac{9p^2p_B}{5} + \frac{6p^2}{5} + \frac{12pp_B}{5} - \frac{8p}{5} - \frac{4p_B}{5} + 1
\end{split}
\end{equation}
\begin{equation}
\begin{split}
\langle F_{BF, BF, PF}\rangle = -\frac{4p^2p_B}{5} + \frac{6p^2}{5} + \frac{16pp_B}{15} - \frac{8p}{5} - \frac{8p_B}{15} + 1
\end{split}
\end{equation}
\begin{equation}
\begin{split}
\langle F_{BF, BF, DP}\rangle = -\frac{27p^2p_B}{20} + \frac{6p^2}{5} + \frac{9pp_B}{5} - \frac{8p}{5} - \frac{3p_B}{5} + 1
\end{split}
\end{equation}
\begin{equation}
\begin{split}
\langle F_{BF, BF, AD}\rangle = -\frac{4p^2p_B}{5} + \frac{2p^2\sqrt{1-p_B}}{15} + \frac{4p^2}{5} + \frac{16pp_B}{15} - \frac{8p\sqrt{1-p_B}}{15} - \\ \frac{16p}{15} -\frac{2p_B}{5} + \frac{4\sqrt{1-p_B}}{15} +\frac{11}{15}
\end{split}
\end{equation}

\begin{figure}[ht!]
\centering
    \begin{subfigure}[b]{0.3\textwidth}
        \includegraphics[width=\textwidth]{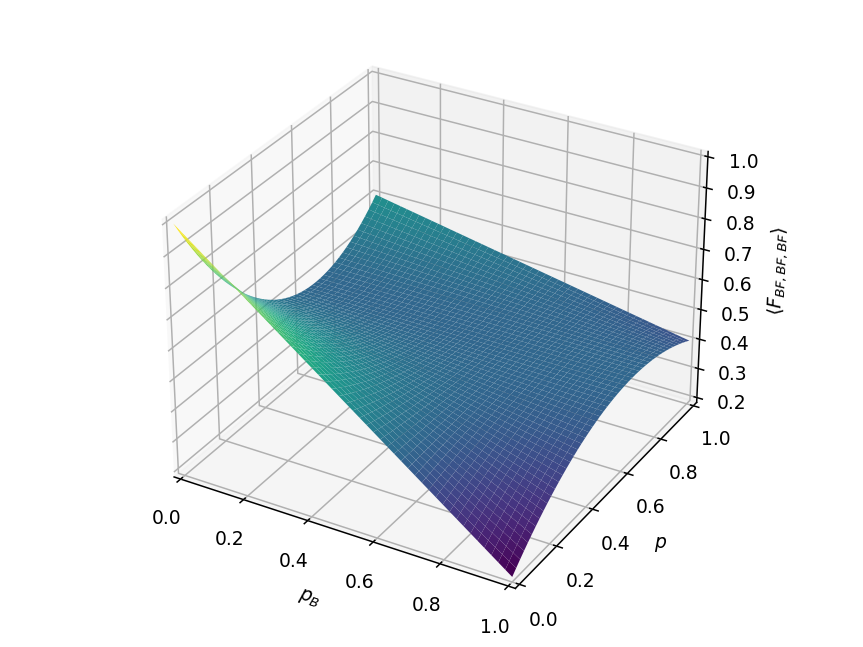}
        \caption{$\langle F_{BF,BF,BF}\rangle$ as a function of $ p$ and $p_B$}
        \label{fig7a}
    \end{subfigure}
    \hfill
    \begin{subfigure}[b]{0.3\textwidth}
        \includegraphics[width=\textwidth]{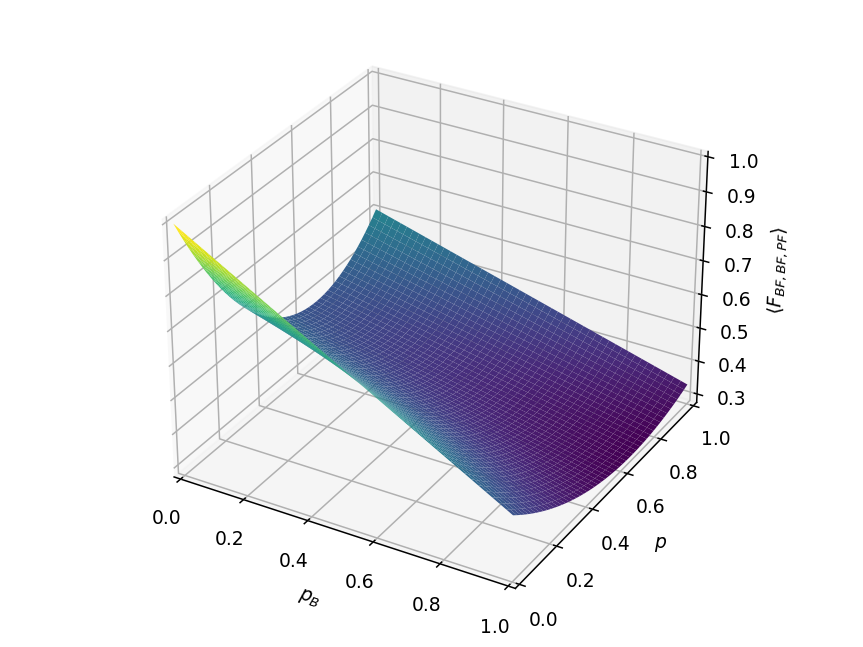}
        \caption{$\langle F_{BF,BF,PF}\rangle$ as a function of $p$ and $p_B$}
        \label{fig7b}
    \end{subfigure}
    \hfill
    \begin{subfigure}[b]{0.3\textwidth}
        \includegraphics[width=\textwidth]{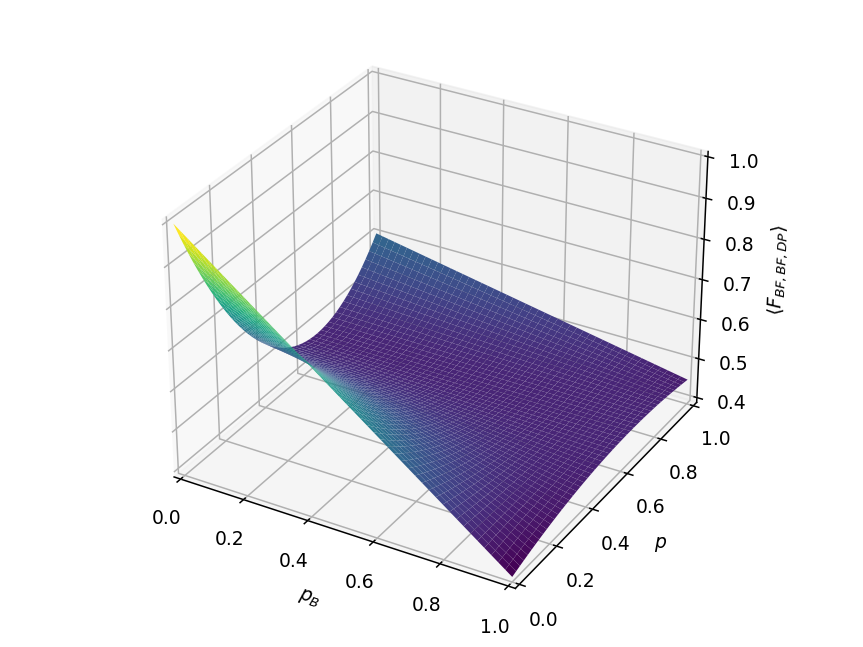}
        \caption{$\langle F_{BF,BF,DP}\rangle$ as a function of $p$ and $p_B$}
        \label{fig7c}
    \end{subfigure}
    \hfill
    \begin{subfigure}[b]{0.3\textwidth}
        \includegraphics[width=\textwidth]{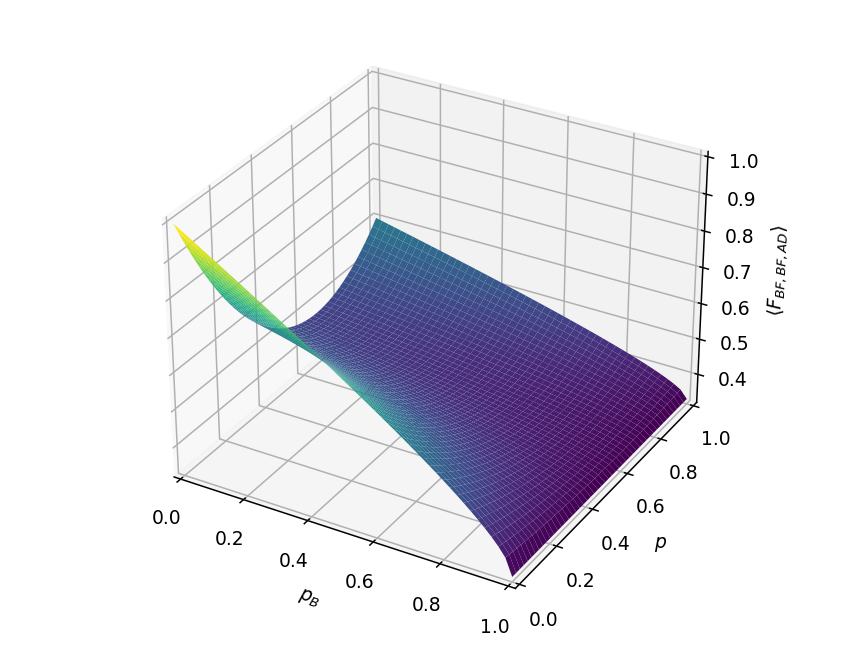}
        \caption{$\langle F_{BF,BF,AD}\rangle$ as a function of $p$ and $p_B$ }
        \label{fig7d}
    \end{subfigure}
    \hfill
    \begin{subfigure}[b]{0.3\textwidth}
        \includegraphics[width=\textwidth]{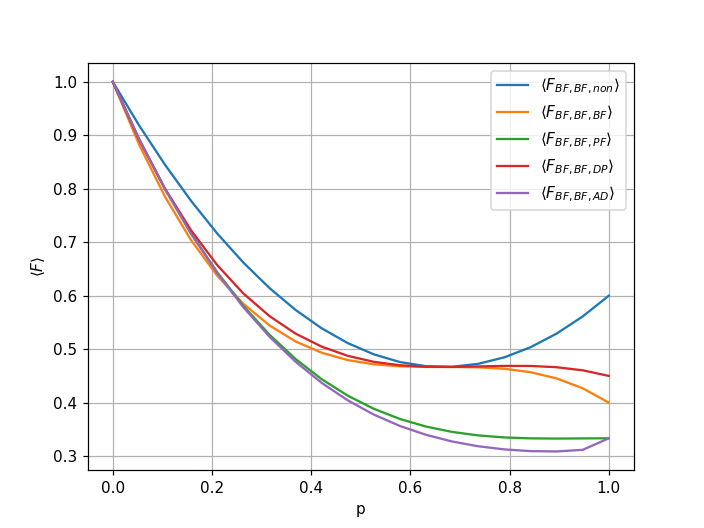}
        \caption{Comparison of FIG 7(a), FIG 7(b), FIG 7(c), FIG 7(d) as $p = p_B = p$}
        \label{fig7e}
    \end{subfigure}
\caption{Constant Bit Flip acting on Alice input state and channel state and Different noise acting on Bob's state}
\label{figure7}
\end{figure}

FIG.(\ref{figure7}), constant bit flip noise acts in the Alice both states with probability $p$ and different noise in channel state $p_B$.
FIG.(\ref{fig7d}) shows a minimum constant fidelity ($\frac{1}{3}$) independent of Amplitude Damping at Bob's end ($p_B$) when Bit Flip is at maximum on both Alice stat($p=1$). In FIG.(\ref{fig7e}), $\langle F_{BF,BF,PF}\rangle = \langle F_{BF,BF,AD}\rangle < \langle F_{BF,BF,BF}\rangle < \langle F_{BF,BF,DP}\rangle < \langle F_{BF,BF,non}\rangle$.

\subsection{Constant Phase flip acting on Alice input state and channel state and Different noise acting on Bob's state}

Let us consider what happens when a constant Phase flip  acts on Alice input state and Alice channel state with probabilities $p_I$ and $p_A$  respectively and different types of noise act on Bob's channel state with probability $p_B$. The fidelity in each case can be written as

\begin{equation}
\begin{split}
\langle F_{PF, PF, BF}\rangle = -\frac{32p^2p_B}{45} + \frac{32p^2}{45} + \frac{16pp_B}{15} - \frac{16p}{15} - \frac{4p_B}{5} + 1
\end{split}
\end{equation}

\begin{equation}
\begin{split}
\langle F_{PF, PF, PF}\rangle = -\frac{16p^2p_B}{15} + \frac{32p^2}{45} + \frac{68pp_B}{45}  -  \frac{16p}{15} - \ \frac{8p_B}{15} + 1
\end{split}
\end{equation}

\begin{equation}
\begin{split}
\langle F_{PF, PF, DP}\rangle = -\frac{8p^2p_B}{15} + \frac{32p^2}{45} + \frac{4pp_B}{5} - \frac{16p}{15} - \frac{3p_B}{5} + 1
\end{split}
\end{equation}

\begin{equation}
\begin{split}
\langle F_{PF, PF, AD}\rangle =  -\frac{16p^2p_B}{45} + \frac{16p^2\sqrt{1-p_B}}{45} + \frac{16p^2}{45} + \frac{4pp_B}{9} + \frac{28p\sqrt{1-p_B}}{45}\\ - \frac{4p}{9} - \frac{2p_B}{5} + \frac{4\sqrt{1-p_B}}{15} + \frac{11}{15}
\end{split}
\end{equation}

\begin{figure}[ht!]
\centering
    \begin{subfigure}[b]{0.3\textwidth}
        \includegraphics[width=\textwidth]{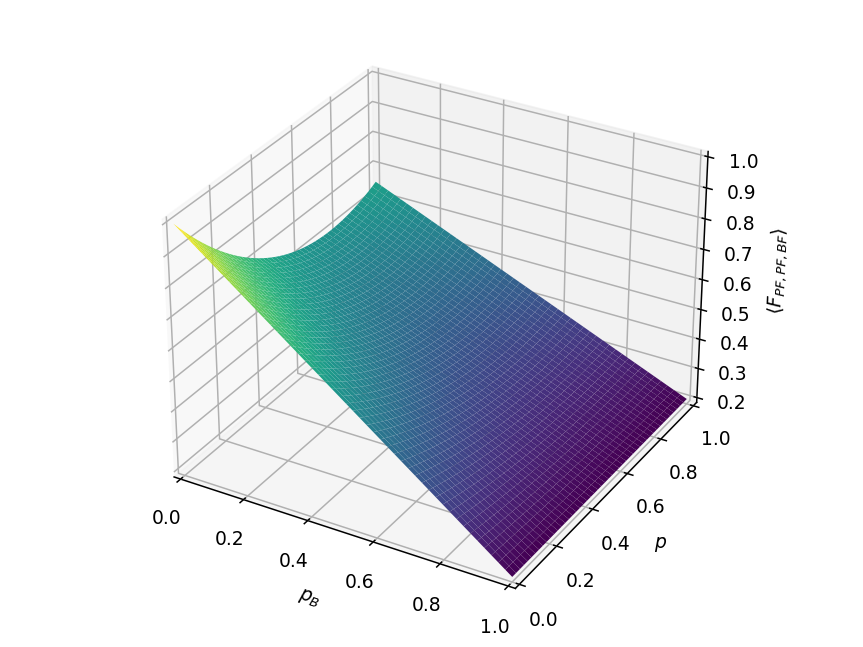 }
        \caption{$\langle F_{PF,PF,BF}\rangle$ as a function of $ p$ and $p_B$ }
    \end{subfigure}
    \hfill
    \begin{subfigure}[b]{0.3\textwidth}
        \includegraphics[width=\textwidth]{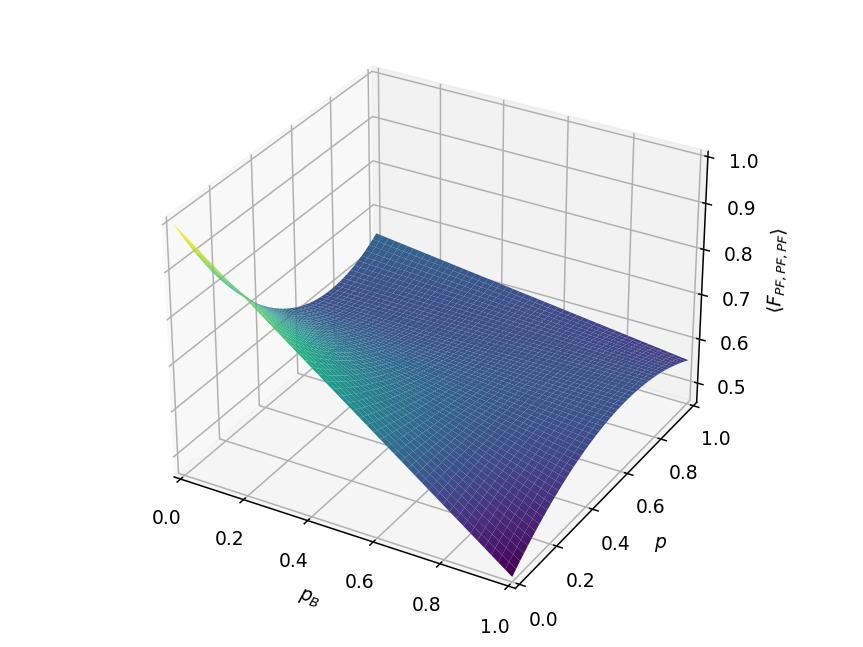  }
        \caption{$\langle F_{PF,PF,PF}\rangle$ as a function of $ p$ and $p_B$}
    \end{subfigure}
    \hfill
    \begin{subfigure}[b]{0.3\textwidth}
        \includegraphics[width=\textwidth]{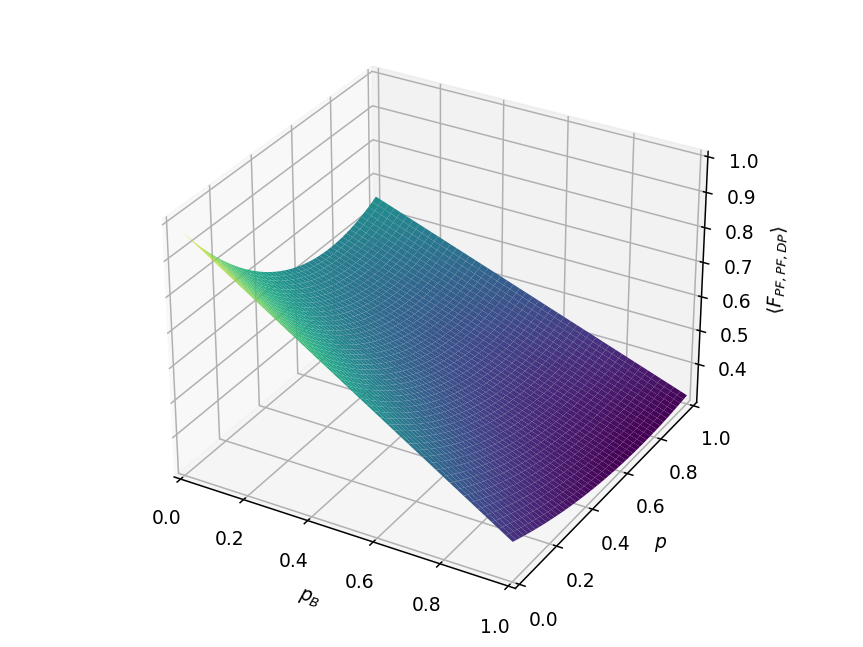 }
        \caption{$\langle F_{PF,PF,DP}\rangle$ as a function of $ p$ and $p_B$}
    \end{subfigure}
    \hfill
    \begin{subfigure}[b]{0.3\textwidth}
        \includegraphics[width=\textwidth]{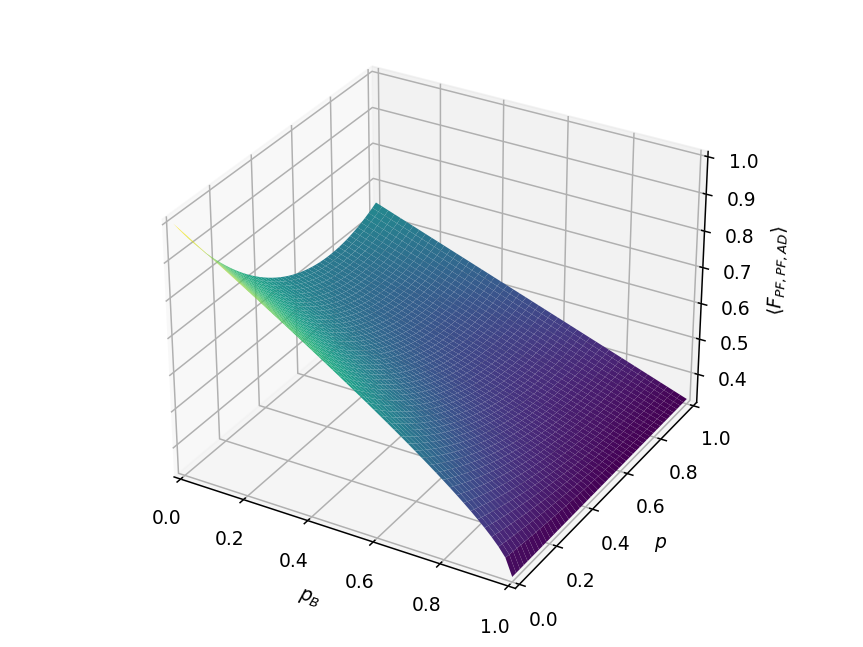}
        \caption{$\langle F_{PF,PF,AD}\rangle$ as a function of $ p$ and $p_B$}
        \label{figure:8d}
    \end{subfigure}
    \hfill
    \begin{subfigure}[b]{0.3\textwidth}
        \includegraphics[width=\textwidth]{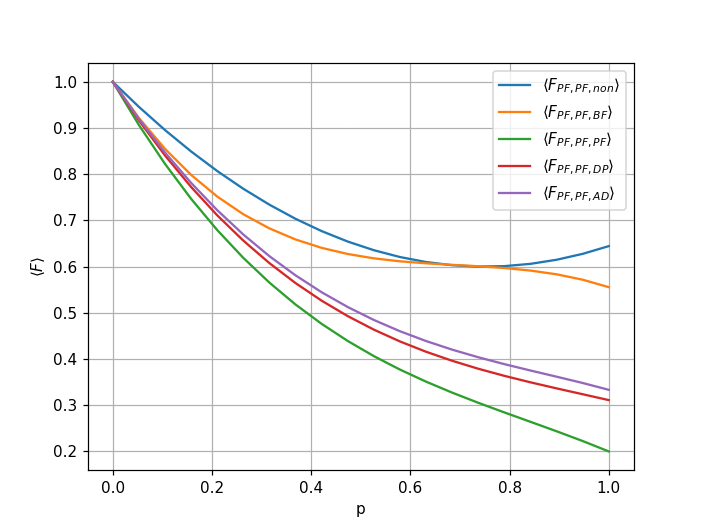}
        \caption{Comparison of FIG 8(a), FIG 8(b), FIG 8(c), FIG 8(d) as $p = p_B = p$}
        \label{figure:8e}
    \end{subfigure}
\caption{Constant Phase flip acting on Alice input state and channel state and Different noise acting on Bob's state}
\label{figure8}
\end{figure}

FIG.(\ref{figure8}) has a constant fidelity ($\frac{1}{5}$) when the noise at Bob's states (Bit flip) are maximum $p_B=1$ and is independent of the input Phase flip noise $p$. FIG.(\ref{figure:8d}) shows a minimum constant fidelity ($\frac{1}{3}$) independent of Phase Flip at Alice both state ($p$) when Amplitude Damping is  maximum on Bob's states($p_B=1$). In FIG.(\ref{figure:8e}) $\langle F_{PF,PF,PF}\rangle < \langle F_{PF,PF,DP}\rangle < \langle F_{PF,PF,AD}\rangle < \langle F_{PF,PF,BF}\rangle < \langle F_{PF,PF,non}\rangle$.

\subsection{Constant Depolarizing noise  acting on Alice input state and channel state and Different noise acting on Bob's state}
 
 We have already gone through how the constant bit flip and phase flip act on  Alice's input state and her entangled channel state when different noises act on Bob's state. Now let us consider the action of depolarizing noise acting on Alice input and channel state and different noises in Bob's state. The average fidelity in the various cases mentioned above can be written as 

\begin{equation}
\begin{split}
\langle F_{DP, DP, BF}\rangle = -\frac{81p^2p_B}{80} + \frac{27p^2}{40} + \frac{9pp_B}{5} - \frac{6p}{5} - \frac{4p_B}{5} + 1
\end{split}
\label{eq:56}
\end{equation}
\begin{equation}
\begin{split}
\langle F_{DP, DP, PF}\rangle = -\frac{9p^2p_B}{20} + \frac{27p^2}{40} + \frac{4pp_B}{5} - \frac{6p}{5} - \frac{8p_B}{15} + 1
\end{split}
\end{equation}
\begin{equation}
\begin{split}
\langle F_{DP, DP, DP}\rangle = -\frac{243p^2p_B}{320} + \frac{27p^2}{40} + \frac{27pp_B}{20}  -  \frac{6p}{5} -  \frac{3p_B}{5} + 1
\end{split}
\end{equation}
\begin{equation}
\begin{split}
\langle F_{DP, DP, AD}\rangle = -\frac{9p^2p_B}{20} + \frac{9p^2\sqrt{1-p_B}}{40} + \frac{9p^2}{20}  + \frac{4pp_B}{5} - \frac{2p\sqrt{1-p_B}}{5}\\ -  \frac{4p}{5} - \frac{2p_B}{5} + \frac{4\sqrt{1-p_B}}{15} + \frac{11}{15}
\end{split}
\label{eq:59}
\end{equation}

\begin{figure}[ht!]
\centering
    \begin{subfigure}[b]{0.3\textwidth}
        \includegraphics[width=\textwidth]{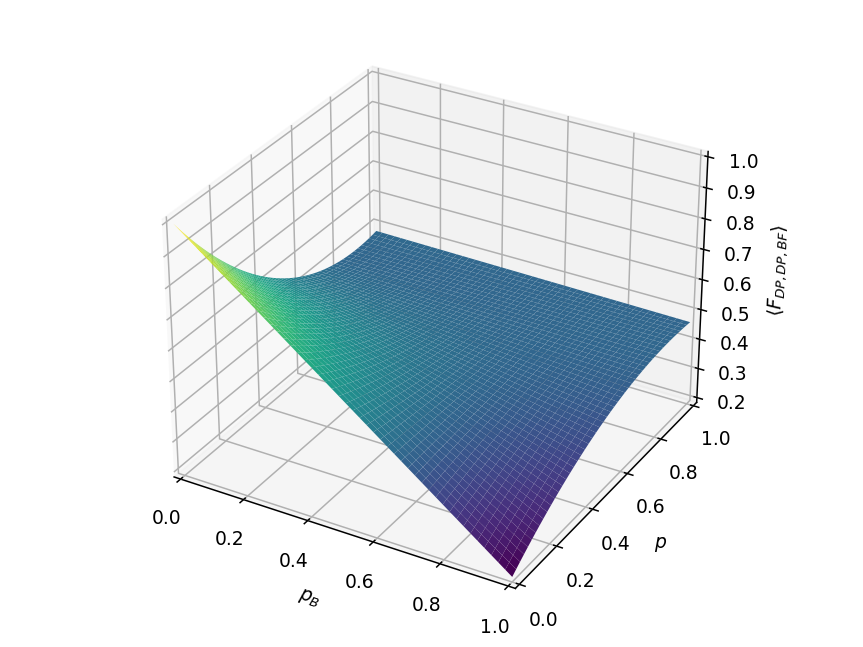 }
        \caption{$\langle F_{DP,DP,BF}\rangle$ as a function of $ p$ and $p_B$}
    \end{subfigure}
    \hfill
    \begin{subfigure}[b]{0.3\textwidth}
        \includegraphics[width=\textwidth]{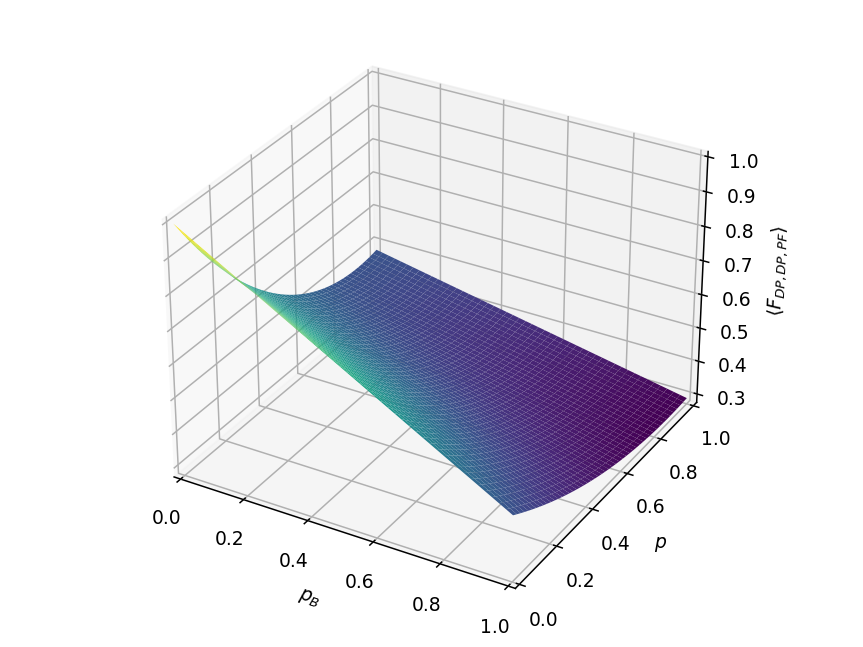}
        \caption{$\langle F_{DP,DP,PF}\rangle$ as a function of $ p$ and $p_B$}
    \end{subfigure}
    \hfill
    \begin{subfigure}[b]{0.3\textwidth}
        \includegraphics[width=\textwidth]{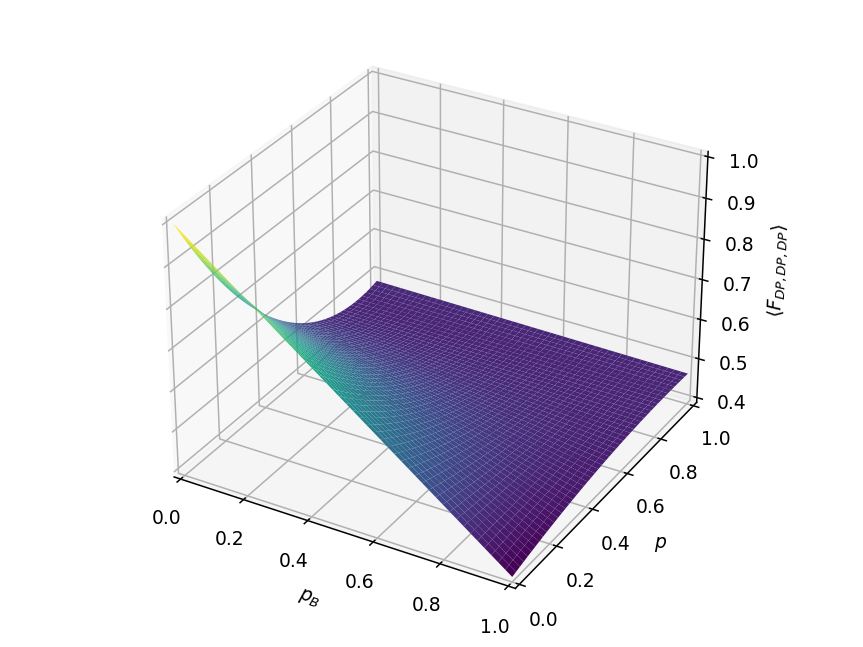} 
        \caption{$\langle F_{DP,DP,DP}\rangle$ as a function of $ p$ and $p_B$}
    \end{subfigure}
    \hfill
    \begin{subfigure}[b]{0.3\textwidth}
        \includegraphics[width=\textwidth]{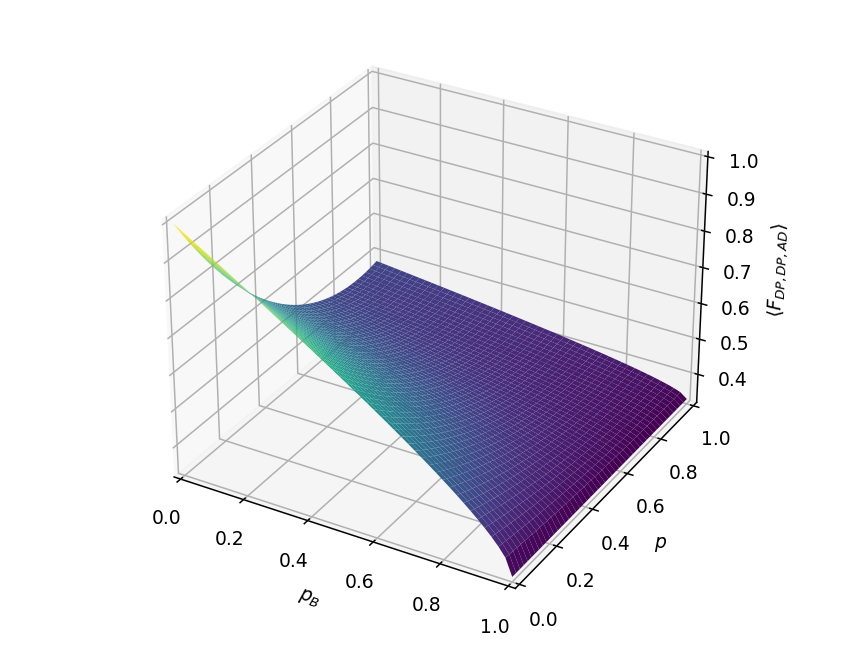}
        \caption{$\langle F_{DP,DP,AD}\rangle$ as a function of $ p$ and $p_B$}
        \label{figu9d}
    \end{subfigure}
    \hfill
    \begin{subfigure}[b]{0.3\textwidth}
        \includegraphics[width=\textwidth]{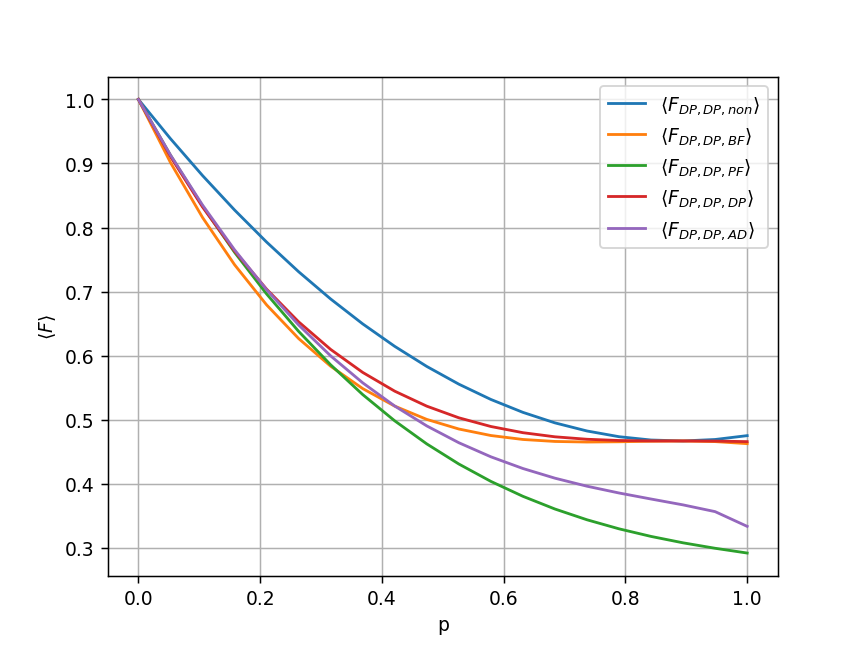}
        \caption{a comparison of FIG 9(a), FIG 9(b), FIG 9(c), FIG 9(d) as $p = p_B = p$ }
        \label{fig9e}
    \end{subfigure}
\caption{Constant Depolarizing noise  acting on Alice input state and channel state and Different noise acting on Bob's state}
\label{fig:figure 9}
\end{figure}
FIG.(\ref{fig:figure 9}) show the fidelity variation described from Eq.(\ref{eq:56}) to Eq.(\ref{eq:59}). In FIG.(\ref{figu9d}) shows a minimum constant fidelity ($\frac{1}{3}$) and independent of Depolarization at both Alice state ($p$) when Amplitude Damping is maximum on Bob's end ($p_B=1$). In FIG.(\ref{fig9e}) $\langle F_{PF,PF,PF}\rangle < \langle F_{PF,PF,DP}\rangle < \langle F_{PF,PF,AD}\rangle < \langle F_{PF,PF,BF}\rangle < \langle F_{PF,PF,non}\rangle$.

\subsection{Constant  Amplitude damping  acting on Alice input state and channel state and Different noise acting on Bob's state}

Let us consider what happens when a constant Amplitude damping acts on Alice input state and Alice channel state with probabilities $p_I$ and $p_A$  respectively and different types of noise act on Bob's channel state with probability $p_B$. The fidelity in each case can be written as,

\begin{equation}
\begin{split}
\langle F_{AD, AD, BF}\rangle = -\frac{4p^2p_B}{9} + \frac{14p^2}{45} + \frac{8pp_B\sqrt{1-p}}{45} + \frac{16pp_B}{15} - \frac{8p\sqrt{1-p}}{45} \\ - \frac{4p}{5} -\frac{8p_B\sqrt{1-p}}{45} -\frac{28p_B}{45} + \frac{8\sqrt{1-p}}{45} + \frac{37}{45}
\end{split}
\label{eq:60}
\end{equation}
\begin{equation}
\begin{split}
\langle F_{AD, AD, PF}\rangle = -\frac{4p^2p_B}{45} + \frac{14p^2}{45} + \frac{4pp_B\sqrt{1-p}}{15} + \frac{16pp_B}{45} - \frac{8p\sqrt{1-p}}{45} - \\  \frac{4p}{5} -\frac{4p_B\sqrt{1-p}}{15} -\frac{4p_B}{15} + \frac{8\sqrt{1-p}}{45} + \frac{37}{45}
\end{split}
\label{eq:61}
\end{equation}
\begin{equation}
\begin{split}
\langle F_{AD, AD, DP}\rangle = -\frac{p^2p_B}{3} + \frac{14p^2}{45} + \frac{2pp_B\sqrt{1-p}}{15} + \frac{4pp_B}{5} - \frac{8p\sqrt{1-p}}{45} - \\  \frac{4p}{5} -\frac{2p_B\sqrt{1-p}}{15} -\frac{7p_B}{15} + \frac{8\sqrt{1-p}}{45} + \frac{37}{45}
\end{split}
\label{eq:62}
\end{equation}
\begin{equation}
\begin{split} 
\langle F_{AD, AD, AD}\rangle = -\frac{26p^2p_B}{45} + \frac{14p^2}{45} + \frac{4pp_B\sqrt{1-p}}{45}  +  \frac{8pp_B}{9} - \frac{4p\sqrt{1-p}\sqrt{1-p_B}}{45} -\\ \frac{4p\sqrt{1-p}}{45} -   \frac{28p}{45}  - \frac{4p_B(1-p)}{45} - \frac{14p_B}{45} + \frac{4\sqrt{1-p}\sqrt{1-p_B}}{45}+ \\\frac{8(1-p)\sqrt{1-p_B}}{45} + \frac{4\sqrt{1-p}}{45} +\frac{29}{45}
\end{split}
\label{eq:63}
\end{equation}

\begin{figure}[ht!]
\centering
    \begin{subfigure}[b]{0.3\textwidth}
        \includegraphics[width=\textwidth]{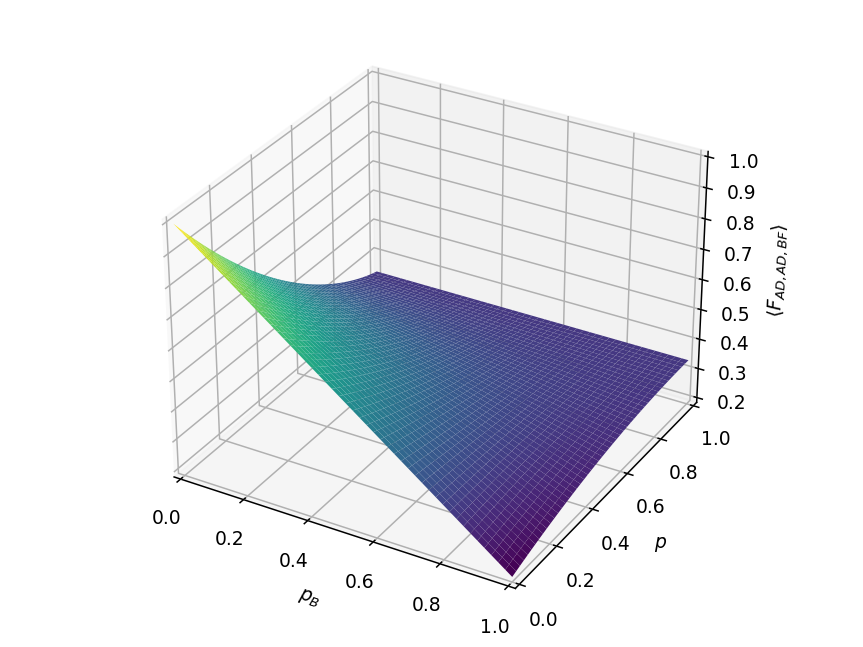 }
        \caption{$\langle F_{AD,AD,BF}\rangle$ as a function of $ p$ and $p_B$}
        \label{fig:10a}
    \end{subfigure}
    \hfill
    \begin{subfigure}[b]{0.3\textwidth}
        \includegraphics[width=\textwidth]{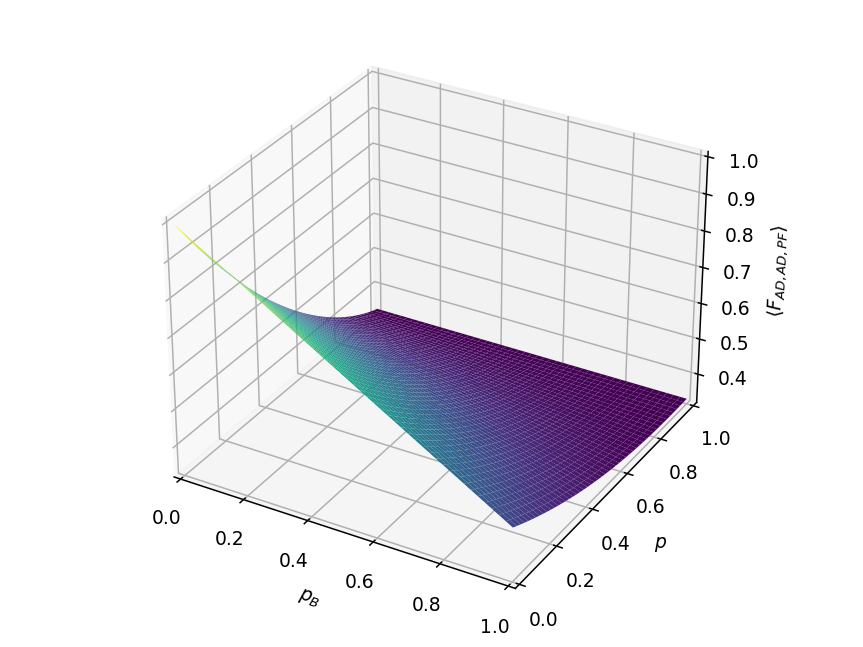}
        \caption{$\langle F_{AD,AD,PF}\rangle$ as a function of $ p$ and $p_B$}
        \label{fig:10b}
    \end{subfigure}
    \hfill
    \begin{subfigure}[b]{0.3\textwidth}
        \includegraphics[width=\textwidth]{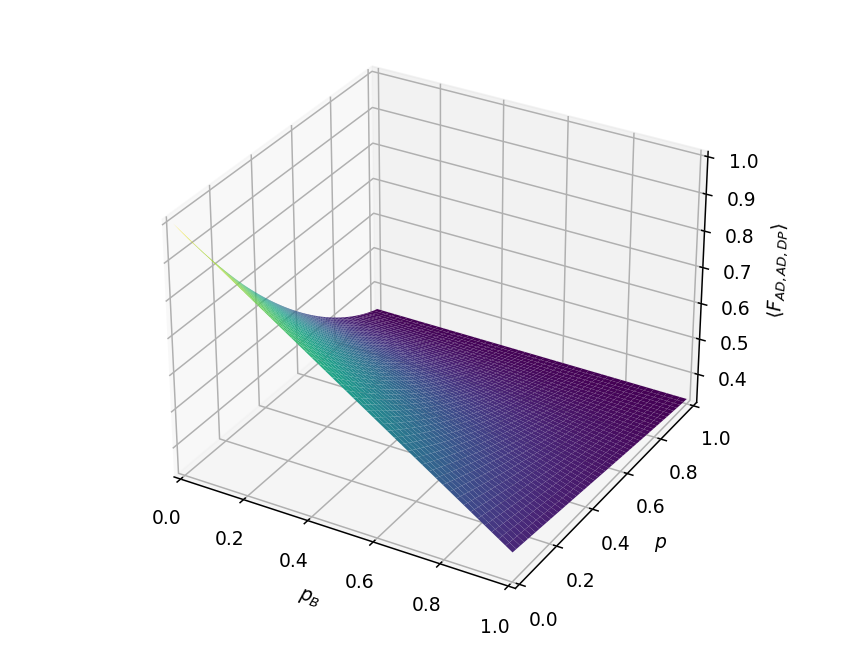} 
        \caption{$\langle F_{AD,AD,DP}\rangle$ as a function of $ p$ and $p_B$}
        \label{fig:10c}
    \end{subfigure}
    \hfill
    \begin{subfigure}[b]{0.3\textwidth}
        \includegraphics[width=\textwidth]{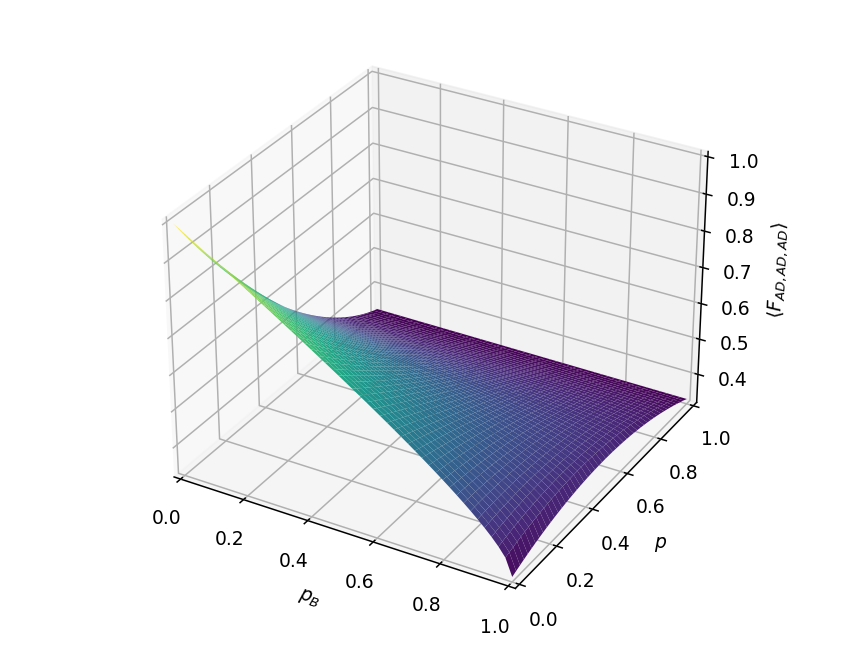}
        \caption{$\langle F_{AD,AD,AD}\rangle$ as a function of $ p$ and $p_B$}
        \label{fig:10d}
    \end{subfigure}
    \hfill
    \begin{subfigure}[b]{0.3\textwidth}
        \includegraphics[width=\textwidth]{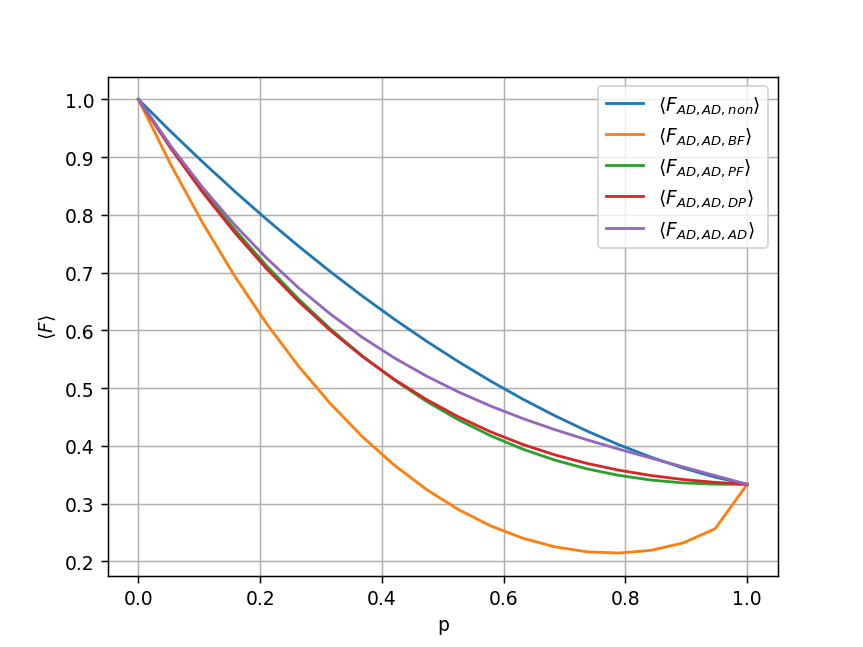}
        \caption{a comparison of FIG 10(a), FIG 10(b), FIG 10(c), FIG 10(d) as $p = p_B = p$ }
        \label{fig:10e}
    \end{subfigure}
\caption{Constant  Amplitude damping  acting on Alice input state and channel state and Different noise acting on Bob's state}
\label{fig:10}
\end{figure}

\vspace{10pt}
The Eq.(\ref{eq:60} to \ref{eq:63}) describing the case considering above showed 3-dimensionally in FIG.(\ref{fig:10a} to \ref{fig:10d}).
FIG.(\ref{fig:10e}) illustrates an extremely intriguing result, when amplitude damping is acting on both Alice states, we obtain constant fidelity ($\frac{1}{3}$) regardless of the noise present in Bob's channel state. If all noise is at its maximum, $p = p_B = 1$.

\section{Constant noise acting in channel state }\label{sec:Constant noise acting in channel state}
Now we want to study the case when same noise act on channel state and different noise acts on the Alice input state.  
\subsection{Noise acting on Alice and Bob's channel state}

Let us assume that Alice's input state is totally isolated from all other types of noise, or that $p_I = 0$, we will now analyse the case where a constant noise acts on the teleportation channel with the same probability $p_A =p_B = p$.

\begin{equation}
\langle F_{non, BF, BF}\rangle = \frac{6p^2}{5} - \frac{8p}{5}  + 1
\label{eq:64}
\end{equation}
\begin{equation}
\langle F_{non, PF, PF}\rangle = \frac{4p^2}{5} - \frac{16p}{15}  + 1
\label{eq:65}
\end{equation}
\begin{equation}
\langle F_{non, DP, DP}\rangle = \frac{27p^2}{40} - \frac{6p}{5} +1
\label{eq:66}
\end{equation}
\begin{equation}
\langle F_{non, AD, AD}\rangle = \frac{2p^2}{3} - \frac{16p}{15} + 1
\label{eq:67}
\end{equation}

\begin{figure}[ht!]
    \centering
    \includegraphics[width=0.5\linewidth]{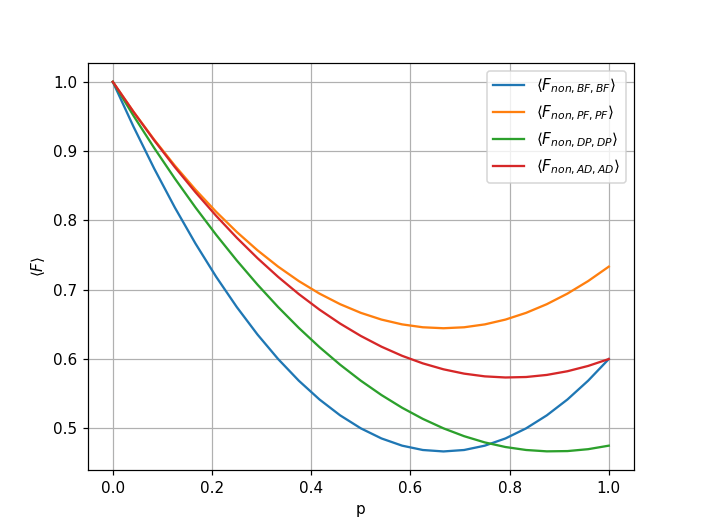}
    \caption{Fidelity v/s probability curve .Comparison of fidelity when constant noise acts on the channel states ($p_A = p_B = p)$ and input state isolated from noise}
\label{fig11}
\end{figure}

FIG.(\ref{fig11}) shows very different conclusion from previous one, in which our system behavior differently such a way that it showing high fidelity compare to when noise acting on any Alice state.Here $\langle F_{non,DP,DP}\rangle < \langle F_{non,AD,AD}\rangle = \langle F_{non,BF,BF}\rangle < \langle F_{non,PF,PF}\rangle $.

\subsection{Constant Bit Flip acting on channel states and different noise acting on Alice input state}

Now we consider the case of fidelity when constant Bit Flip acting on teleportation channel with same probability $p_A = p_B = p$ and different noise act on Alice input state with probability $p_I$. In each case the fidelity can be written as,

\begin{equation}
\langle F_{BF, BF, BF}\rangle = -\frac{9p^2p_I}{5} + \frac{6p^2}{5} + \frac{12pp_I}{5} - \frac{8p}{5} - \frac{4p_I}{5} + 1
\label{eq:68}
\end{equation}
\begin{equation}
\langle F_{PF, BF, BF}\rangle = -\frac{4p^2p_I}{5} + \frac{6p^2}{5} + \frac{16pp_I}{15} - \frac{8p}{5} - \frac{8p_I}{15} + 1
\label{eq:69}
\end{equation}
\begin{equation}
\begin{split}
\langle F_{DP, BF, BF}\rangle = -\frac{27p^2p_I}{20} + \frac{6p^2}{5} + \frac{9pp_I}{5} - \frac{8p}{5} - \frac{3p_I}{5} + 1
\end{split}
\label{eq:70}
\end{equation}
\begin{equation}
    \begin{split}
\langle F_{AD, BF, BF}\rangle = -\frac{4p^2p_I}{5} + \frac{2p^2\sqrt{1-p_I}}{5} + \frac{4p^2}{5} + \frac{16pp_I}{15} - \frac{8p\sqrt{1-p_I}}{15} - \\ \frac{16p}{15} -  \frac{2p_I}{5} + \frac{4\sqrt{1-p_I}}{15} + \frac{11}{15}
    \end{split}
\label{eq:71}
\end{equation}
\begin{figure}[ht!]
\centering
    \begin{subfigure}[b]{0.3\textwidth}
        \includegraphics[width=\textwidth]{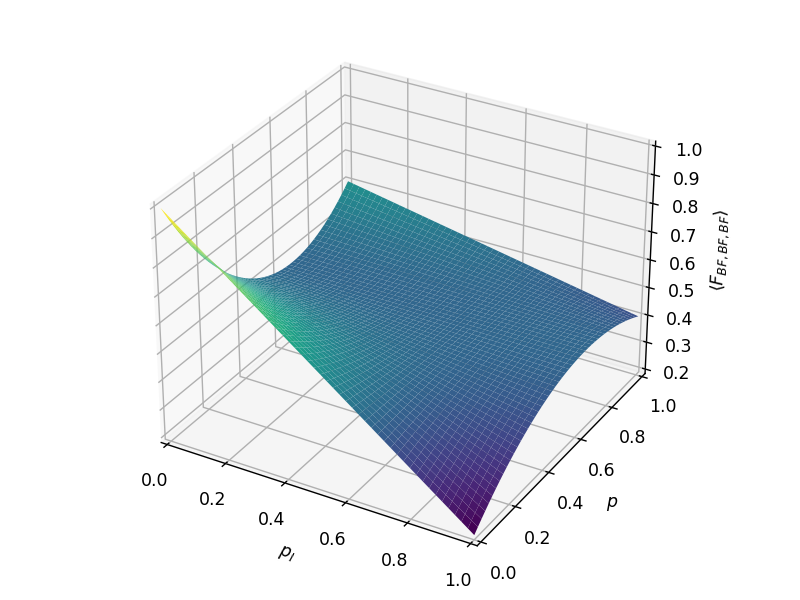}
        \caption{$\langle F_{BF,BF,BF} \rangle$ as a function of $ p_I$ and $p$}
        \label{fig12 (a)}
    \end{subfigure}
    \hfill
    \begin{subfigure}[b]{0.3\textwidth}
        \includegraphics[width=\textwidth]{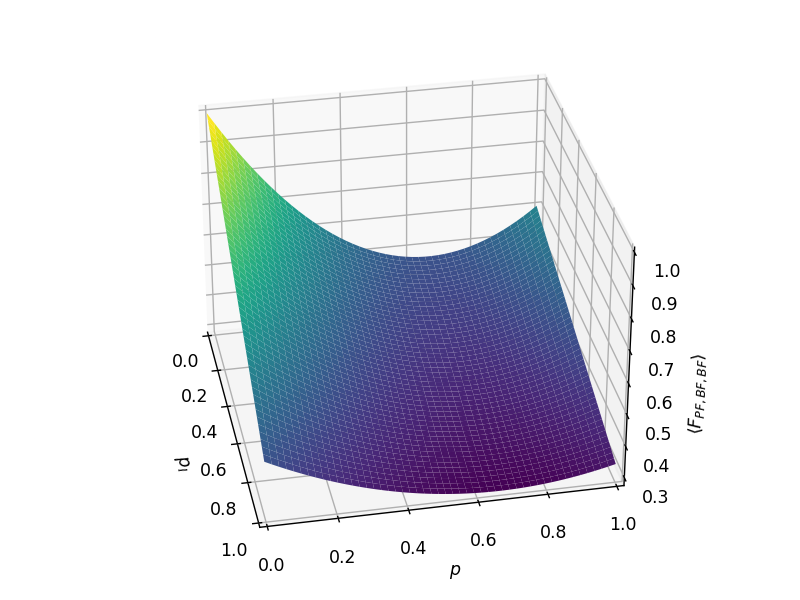}
        \caption{$\langle F_{PF,BF,BF} \rangle$ as a function of $ p_I$ and $p$}
         \label{fig12 (b)}
    \end{subfigure}
    \hfill
    \begin{subfigure}[b]{0.3\textwidth}
        \includegraphics[width=\textwidth]{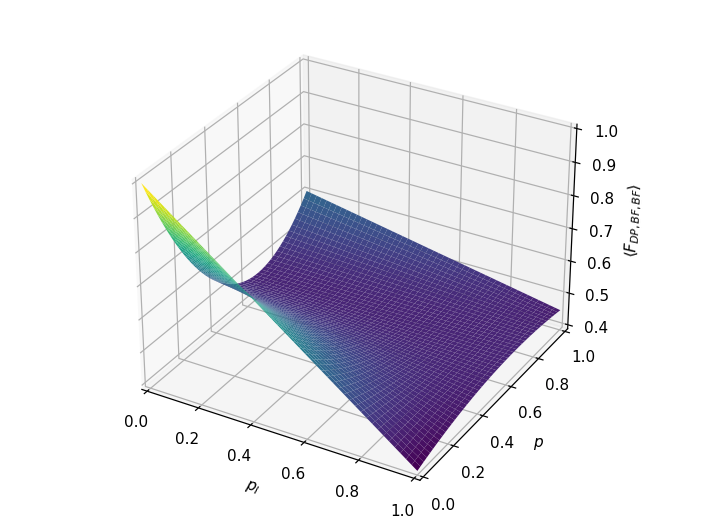}
        \caption{$\langle F_{DP,BF,BF} \rangle$ as a function of $ p_I$ and $p$}
         \label{fig12 (c)}
    \end{subfigure}
    \hfill
    \begin{subfigure}[b]{0.3\textwidth}
        \includegraphics[width=\textwidth]{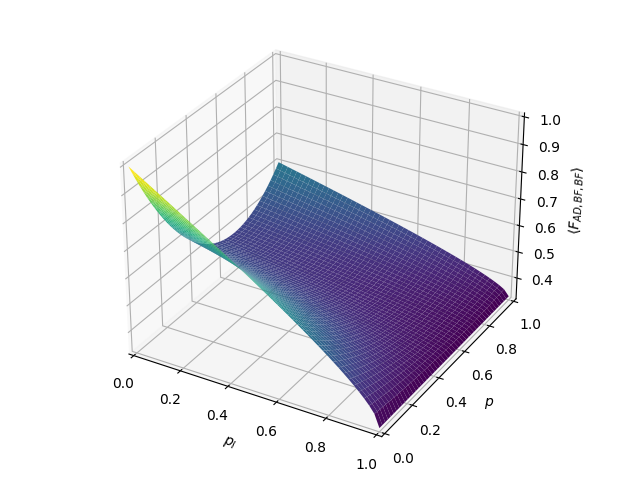}
        \caption{$\langle F_{AD,BF,BF} \rangle$ as a function of $ p_I$ and $p$}
         \label{fig12 (d)}
    \end{subfigure}
    \hfill
    \begin{subfigure}[b]{0.3\textwidth}
        \includegraphics[width=\textwidth]{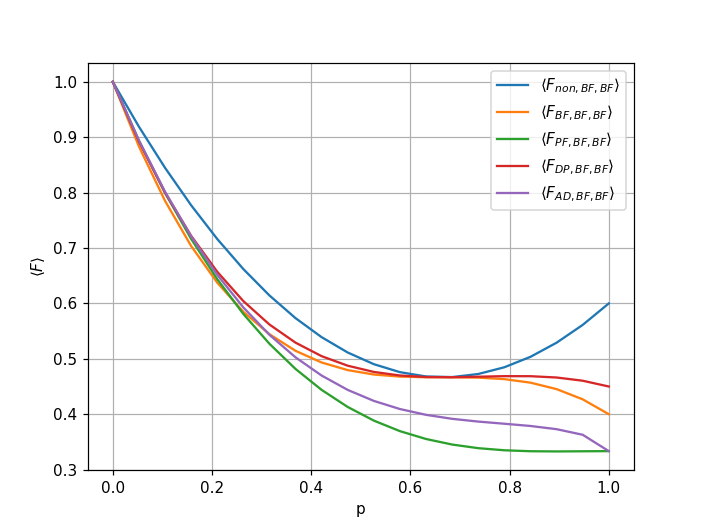}
        \caption{a comparison of FIG 12(a), FIG 12(b), FIG 12(c), FIG 12(d) as $p_I = p = p$ }
         \label{fig12 (e)}
    \end{subfigure}
\caption{Fidelity when constant Bit Flip acting on channel state and different noise acting on Alice input state}
\end{figure}

FIG.(\ref{fig12 (a)}) to FIG.(\ref{fig12 (d)}) shows the variation of fidelity of Eq.(\ref{eq:68} to \ref{eq:71}). In FIG.(\ref{fig12 (d)}), the Fidelity is constant ($\frac{1}{3}$) when the Amplitude damping noise in the input state is maximum ($p_I = 1$) and is independent of the Bit flip noise at both the channel states ($ p $). FIG.(\ref{fig12 (e)}) shows the two dimensional plot of Eq.(\ref{eq:68} to \ref{eq:71}) and Eq.(\ref{eq:64}), in which $\langle F_{PF,BF,BF}\rangle = \langle F_{AD,BF,BF}\rangle < \langle F_{BF,BF,BF}\rangle < \langle F_{DP,BF,BF}\rangle < \langle F_{non,BF,BF}\rangle$.

\subsection{ Constant Phase Flip acting on channel state and different noise acting on Alice input state}
Now we consider the case when constant Phase Flip acting on channel state with probability $p_A = p_B = p$ and different noise act on Alice channel state with probability $p_I$.
\begin{equation}
\begin{split}
\langle F_{BF, PF, PF}\rangle = -\frac{4p^2p_I}{5} + \frac{4p^2}{5} + \frac{16pp_I}{15}  -  \frac{16p}{15} - \ \frac{4p_I}{5} + 1
\label{eq:72}
\end{split}
\end{equation}
\begin{equation}
\begin{split}
\langle F_{PF, PF, PF}\rangle = -\frac{16p^2p_I}{15} + \frac{4p^2}{5} + \frac{64pp_I}{45}  -  \frac{16p}{15} - \ \frac{8p_I}{15} + 1
\end{split}
\label{eq:73}
\end{equation}
\begin{equation}
\begin{split}
\langle F_{DP, PF, PF}\rangle = -\frac{3p^2p_I}{5} + \frac{4p^2}{5} + \frac{4pp_I}{5}  -  \frac{16p}{15} - \frac{3p_I}{5} +  1
\end{split}
\label{eq:74}
\end{equation}
\begin{equation}
\langle F_{AD, PF, PF}\rangle = -\frac{4p^2p_I}{15} + \frac{8p^2\sqrt{1-p_I}}{15} + \frac{4p^2}{15}  +  \frac{16pp_I}{45} -  \frac{32p\sqrt{1-p_I}}{45} -\\ \frac{16p}{45} -\frac{2p_I}{5}+ \frac{4\sqrt{1-p_I}}{15} + \frac{11}{15}
\label{eq:75}
\end{equation}
\begin{figure}[ht!]
\centering
    \begin{subfigure}[b]{0.3\textwidth}
        \includegraphics[width=\textwidth]{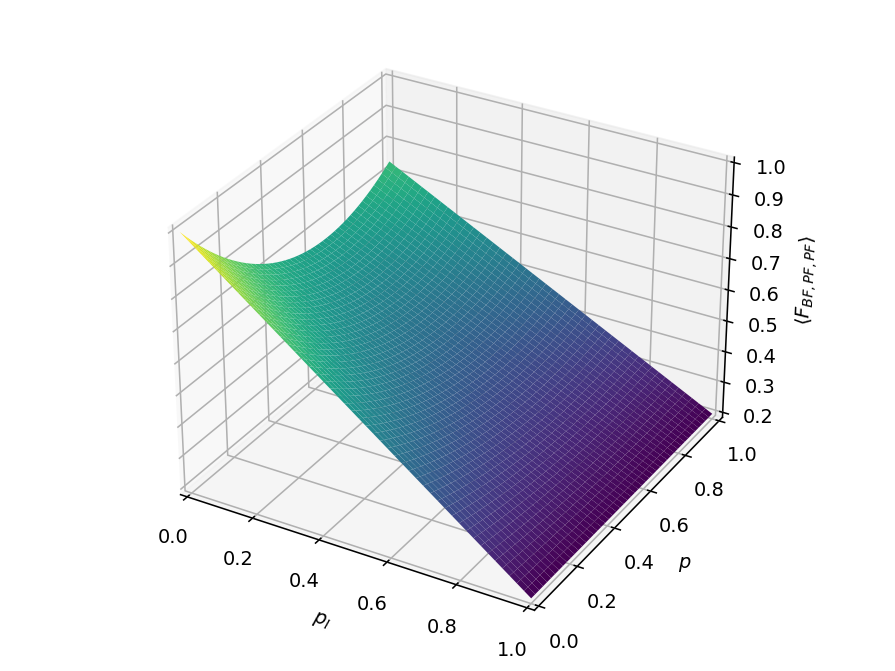}
        \caption{$\langle F_{BF,PF,PF} \rangle$ as a function of $p_I$ and $p$.}
        \label{fig13 (a)}
    \end{subfigure}
    \hfill
    \begin{subfigure}[b]{0.3\textwidth}
        \includegraphics[width=\textwidth]{F_pf_bf_bf.png}
        \caption{$\langle F_{PF,PF,PF} \rangle$ as a function of $p_I$ and $p$.}
        \label{fig13 (b)}
    \end{subfigure}
    \hfill
    \begin{subfigure}[b]{0.3\textwidth}
        \includegraphics[width=\textwidth]{F_dp_bf_bf.png}
        \caption{$\langle F_{DP,PF,PF} \rangle$ as a function of $p_I$ and $p$.}
        \label{fig13 (c)}
    \end{subfigure}
    \hfill
    \begin{subfigure}[b]{0.3\textwidth}
        \includegraphics[width=\textwidth]{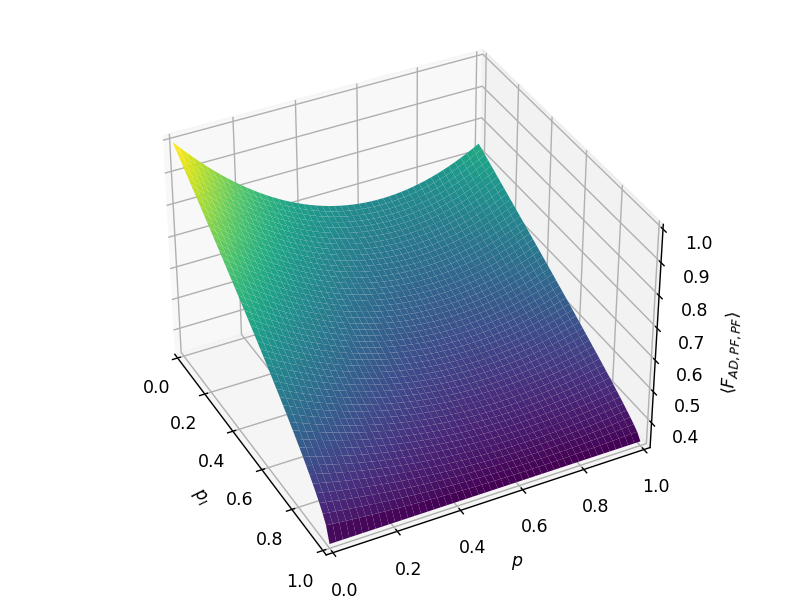}
        \caption{$\langle F_{AD,PF,PF} \rangle$ as a function of $p_I$ and $p$}
        \label{fig13 (d)}
    \end{subfigure}
    \hfill
    \begin{subfigure}[b]{0.3\textwidth}
        \includegraphics[width=\textwidth]{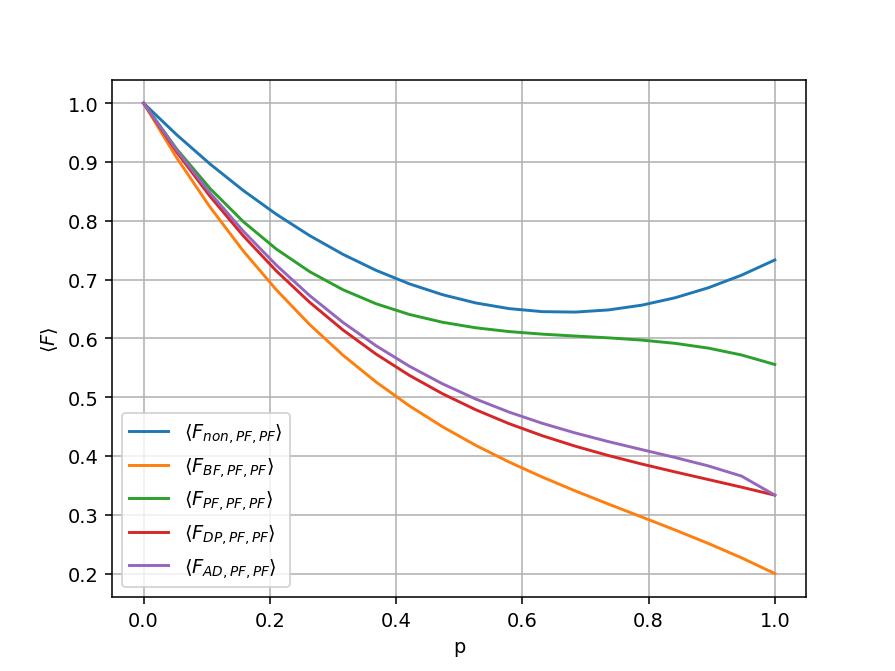}
        \caption{a comparison of FIG 13(a), FIG 13(b), FIG 13(c), FIG 13(d) as $p_I = p = p$ }
        \label{fig13 (e)}
    \end{subfigure}
\caption{Fidelity when constant Phase Flip acting on channel state and different noise acting on Alice input state}
\label{fig:13}
\end{figure}

FIG.(\ref{fig:13}) shows the change in fidelity considered above. In FIG.(\ref{fig13 (a)}), the Fidelity is constant ($\frac{1}{5}$) when the Amplitude damping noise in the input state is maximum ($p_I = 1$) and is independent of the phase flip noise at both the channel states ($ p $). 
In FIG.(\ref{fig13 (d)}), the Fidelity is constant ($\frac{1}{3}$) when the Bit flip noise in the input state is maximum ($p_I = 1$) and is independent of the phase flip noise at both the channel states ($ p $). In FIG.(\ref{fig13 (e)}) shows one dimensional plot of Eq.(\ref{eq:72} to \ref{eq:75}) and Eq.(\ref{eq:65}). From the plot it is observed $\langle F_{BF,PF,PF}\rangle < \langle F_{DP,PF,PF}\rangle = \langle F_{AD,PF,PF}\rangle < \langle F_{PF,PF,PF}\rangle < \langle F_{non,PF,PF}\rangle$.

\subsection{Constant Depolarizing noise acting on channel state and different noise acting on Alice input state}

Now we consider the case when constant Depolarizing noise acting on channel state with probability $p_A = p_B = p$ and different noise act on Alice channel state with probability $p_I$.

\begin{equation}\label{eq:76}
\begin{split}
\langle F_{BF, DP, DP}\rangle = - \frac{81p^2p_I}{80} + \frac{27p^2}{40} + \frac{9pp_I}{5}  -  \frac{6p}{5} - \frac{4p_I}{5} + 1
\end{split}
\end{equation}
\begin{equation}\label{eq:77}
\begin{split}
\langle F_{PF, DP, DP}\rangle = -\frac{9p^2p_I}{20} + \frac{27p^2}{40} + \frac{4pp_I}{5}  -  \frac{6p}{5} - \frac{8p_I}{15} + 1
\end{split}
\end{equation}
\begin{equation}\label{eq:78}
\begin{split}
\langle F_{DP, DP, DP}\rangle = -\frac{243p^2p_I}{320} + \frac{27p^2}{40} + \frac{27pp_I}{20}  -  \frac{6p}{5} -  \frac{3p_I}{5} + 1
\end{split}
\end{equation}
\begin{equation}\label{eq:79}
\begin{split}
\langle F_{AD, DP, DP}\rangle = -\frac{9p^2p_I}{20} + \frac{9p^2\sqrt{1-p_I}}{40} - \frac{2p\sqrt{1-p_I}}{5}  +  \frac{9p^2}{20} +  \frac{4pp_I}{5} - \\ \frac{4p}{5} - \frac{2p_I}{5} + \frac{4\sqrt{1-p_I}}{15} +\frac{11}{15}
\end{split}
\end{equation}

\begin{figure}[ht!]
\centering
    \begin{subfigure}[b]{0.3\textwidth}
        \includegraphics[width=\textwidth]{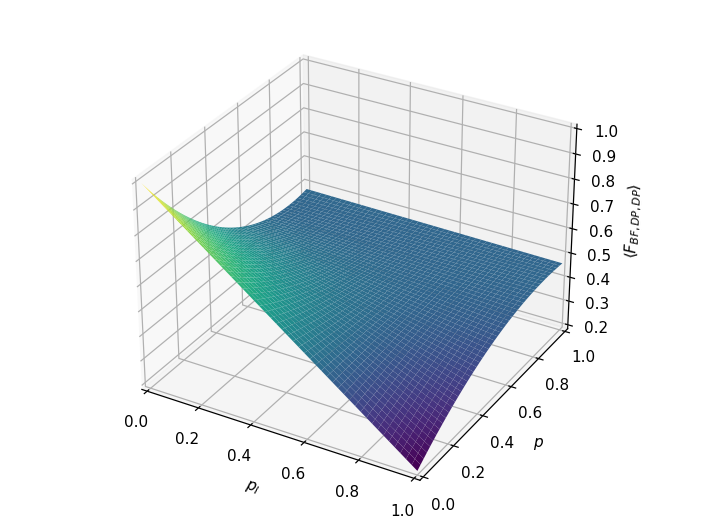}
        \caption{$\langle F_{BF,DP,DP} \rangle$ as a function of $p_I$ and $p$.}
        \label{fig:14(a)}
    \end{subfigure}
    \hfill
    \begin{subfigure}[b]{0.3\textwidth}
        \includegraphics[width=\textwidth]{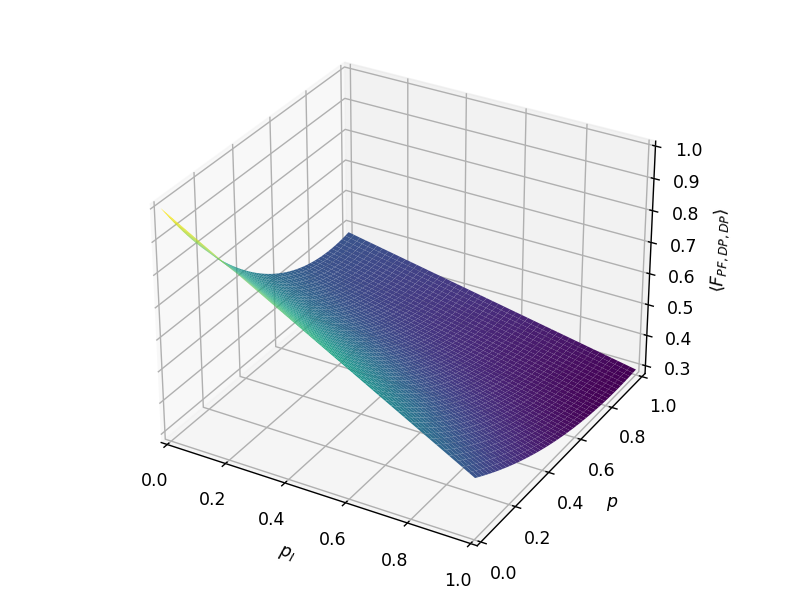}
        \caption{$\langle F_{PF,DP,DP} \rangle$ as a function of $p_I$ and $p$.}
        \label{fig:14(b)}
    \end{subfigure}
    \hfill
    \begin{subfigure}[b]{0.3\textwidth}
        \includegraphics[width=\textwidth]{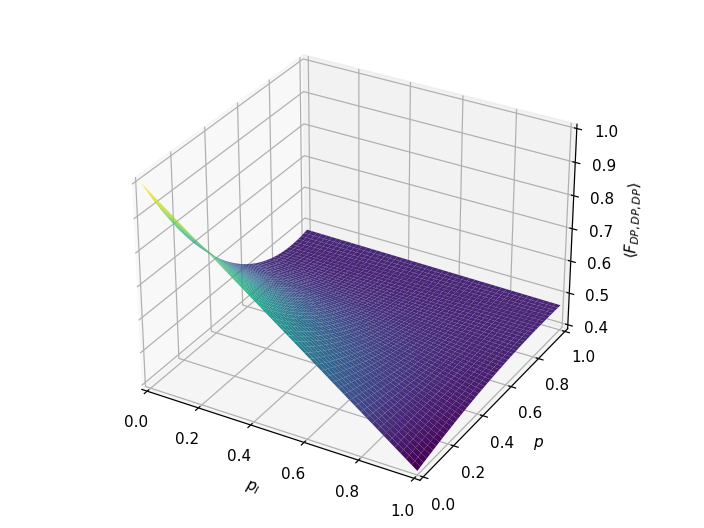}
        \caption{$\langle F_{DP,DP,DP} \rangle$ as a function of $p_I$ and $p$.}
        \label{fig:14(c)}
    \end{subfigure}
    \hfill
    \begin{subfigure}[b]{0.3\textwidth}
        \includegraphics[width=\textwidth]{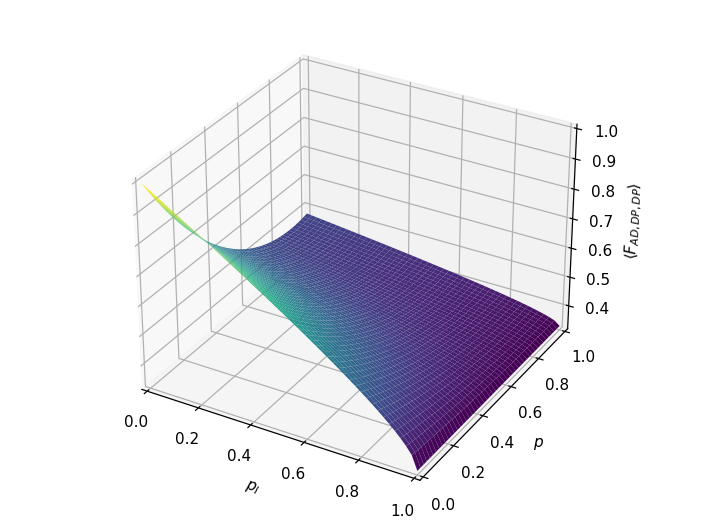}
        \caption{$\langle F_{AD,DP,DP} \rangle$ as a function of $p_I$ and $p$}
        \label{fig:14(d)}
    \end{subfigure}
    \hfill
    \begin{subfigure}[b]{0.3\textwidth}
        \includegraphics[width=\textwidth]{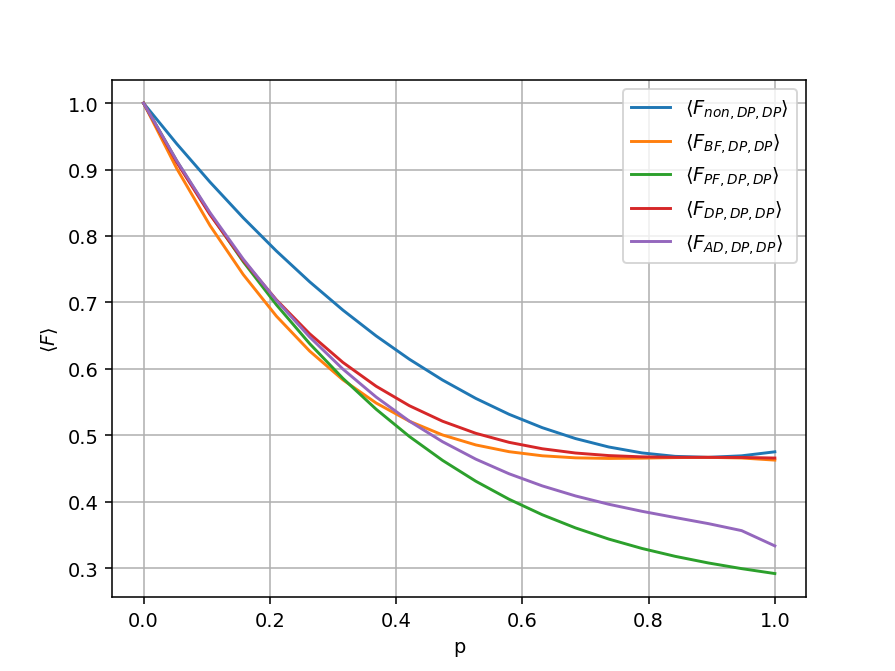}
        \caption{a comparison of FIG 14(a), FIG 14(b), FIG 14(c), FIG 14(d) as $p_I = p = p$ }
        \label{fig:14(e)}
    \end{subfigure}
\caption{Fidelity when constant Depolarizing noise acting on channel state and different noise acting on Alice input state}
\end{figure}

FIG.(\ref{fig:14(a)}) to FIG.(\ref{fig:14(d)}) shows the fidelity variation of Eq.(\ref{eq:76} to \ref{eq:79}).
In FIG.(\ref{fig:14(d)}), the fidelity is constant ($\frac{1}{3}$) which is independent of the depolarizing noise at the channel states ($p$) when the amplitude damping noise at the input is maximum ($p_I=1)$. In FIG.(\ref{fig:14(e)}), we shows a one dimensional comparison of Eq.(\ref{eq:76} to \ref{eq:79}) and Eq.(\ref{eq:66}). In which $\langle F_{PF,DP,DP}\rangle < \langle F_{AD,DP,DP}\rangle < \langle F_{BF,DP,DP}\rangle < \langle F_{DP,DP,DP}\rangle < \langle F_{non,DP,DP}\rangle$.

\subsection{Fidelity when constant Amplitude damping acting on channel state and different noise acting on Alice input state}

Now we consider the case when constant Amplitude damping acting on channel state with probability $p_A = p_B = p$ and different noise act on Alice channel state with probability $p_I$.

\begin{equation}\label{eq:80}
\langle F_{BF, AD, AD}\rangle = -\frac{14p^2p_I}{15} + \frac{2p^2}{3} + \frac{4pp_I}{3}  -  \frac{16p}{15} -  \frac{4p_I}{5} + 1
\end{equation}
\begin{equation}\label{eq:81}
\langle F_{PF, AD, AD}\rangle = -\frac{8p^2p_I}{45} + \frac{2p^2}{3} + \frac{16pp_I}{45}  -  \frac{16p}{15} - \frac{8p_I}{15} + 1
\end{equation}
\begin{equation}\label{eq:82}
\begin{split}
\langle F_{DP, AD, AD}\rangle = -\frac{7p^2p_I}{10} + \frac{2p^2}{3} + pp_I -  \frac{16p}{15} -  \frac{3p_I}{5} +1  
\end{split}
\end{equation}
\begin{equation}\label{eq:83}
\begin{split}
\langle F_{AD, AD, AD}\rangle = -\frac{26p^2p_I}{45} + \frac{4p^2\sqrt{1-p_I}}{45} + \frac{26p^2}{45}  +  \frac{32pp_I}{45}+ -  \frac{16p\sqrt{1-p_I}}{45} -\\ \frac{32p}{45}  - \frac{2p_I}{5} + \frac{4\sqrt{1-p_I}}{15} +\frac{11}{15}
\end{split}
\end{equation}

\begin{figure}[ht!]
\centering
    \begin{subfigure}[b]{0.3\textwidth}
        \includegraphics[width=\textwidth]{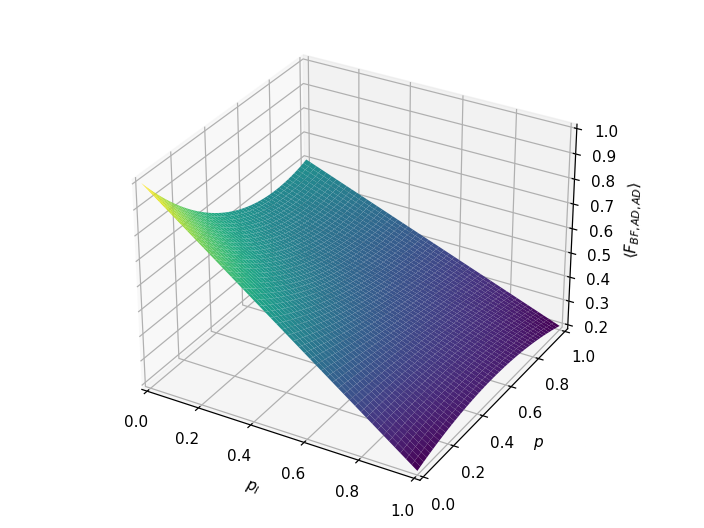}
        \caption{$\langle F_{BF,AD,AD} \rangle$ as a function of $p_I$ and $p$.}
        \label{fig 15(a)}
    \end{subfigure}
    \hfill
    \begin{subfigure}[b]{0.3\textwidth}
        \includegraphics[width=\textwidth]{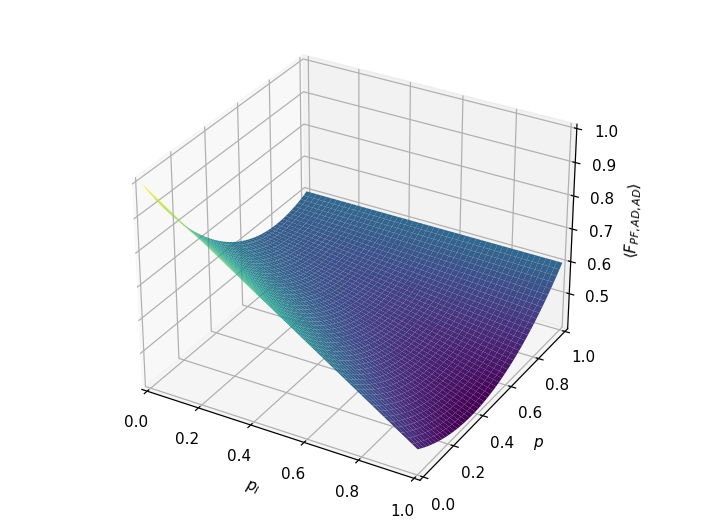}
        \caption{$\langle F_{PF,AD,AD} \rangle$ as a function of $p_I$ and $p$.}
        \label{fig 15(b)}
    \end{subfigure}
    \hfill
    \begin{subfigure}[b]{0.3\textwidth}
        \includegraphics[width=\textwidth]{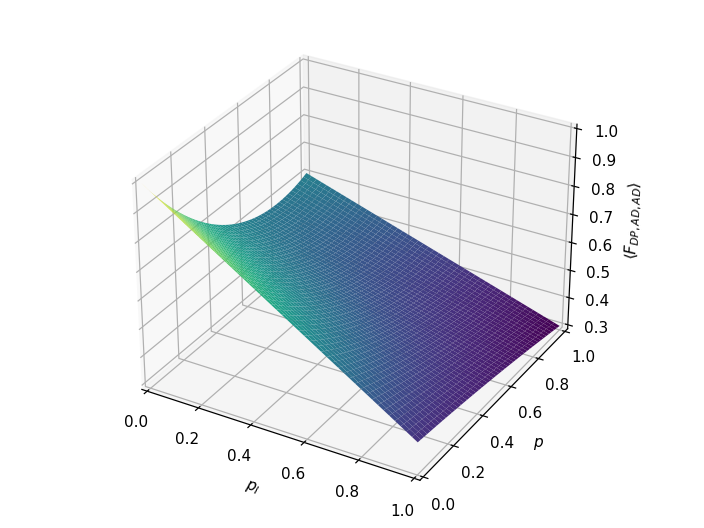}
        \caption{$\langle F_{DP,AD,AD} \rangle$ as a function of $p_I$ and $p$.}
        \label{fig 15(c)}
    \end{subfigure}
    \hfill
    \begin{subfigure}[b]{0.3\textwidth}
        \includegraphics[width=\textwidth]{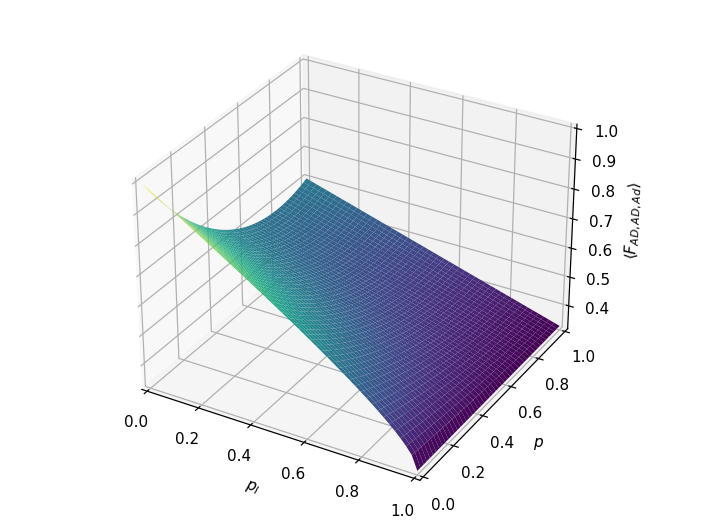}
        \caption{$\langle F_{AD,AD,AD} \rangle$ as a function of $p_I$ and $p$}
        \label{fig 15(d)}
    \end{subfigure}
    \hfill
    \begin{subfigure}[b]{0.3\textwidth}
        \includegraphics[width=\textwidth]{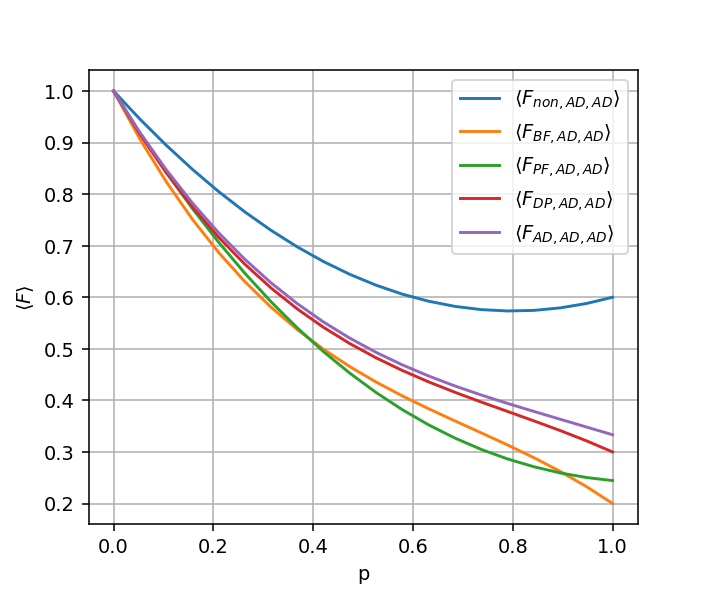}
        \caption{a comparison of FIG 15(a), FIG 15(b), FIG 15(c), FIG 15(d) as $p_I = p = p$ }
    \label{fig 15(e)}
    \end{subfigure}
\caption{Fidelity when constant Amplitude damping acting on channel state and different noise acting on Alice input state.}
\end{figure}

In FIG.(\ref{fig 15(d)}), the fidelity is constant ($\frac{1}{3}$) when the input state has maximum Amplitude damping noise($p_I =1 $) and  both the channel states also have amplitude damping noise which does not affect the fidelity. In FIG.(\ref{fig 15(e)}) we shows a one dimensional comparison of Eq.(\ref{eq:80} to \ref{eq:83}) and Eq.(\ref{eq:67}). In which $\langle F_{BF,AD,AD}\rangle < \langle F_{PF,AD,AD}\rangle < \langle F_{DP,AD,AD}\rangle < \langle F_{AD,AD,AD}\rangle < \langle F_{non,AD,AD}\rangle$.

\subsection{Noise Abstraction}

\begin{figure}[ht!] 
    \begin{subfigure}[b]{0.5\linewidth}
    \centering
        \includegraphics[width=0.5\linewidth]{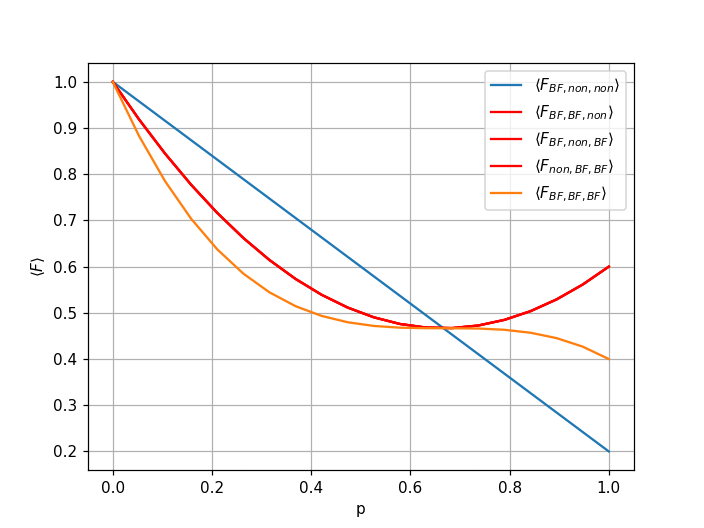} 
        \caption{Bit Flip Noise} 
        \label{fig 16(a)} 
        \vspace{2ex}
    \end{subfigure}
    \begin{subfigure}[b]{0.5\linewidth}
    \centering
        \includegraphics[width=0.5\linewidth]{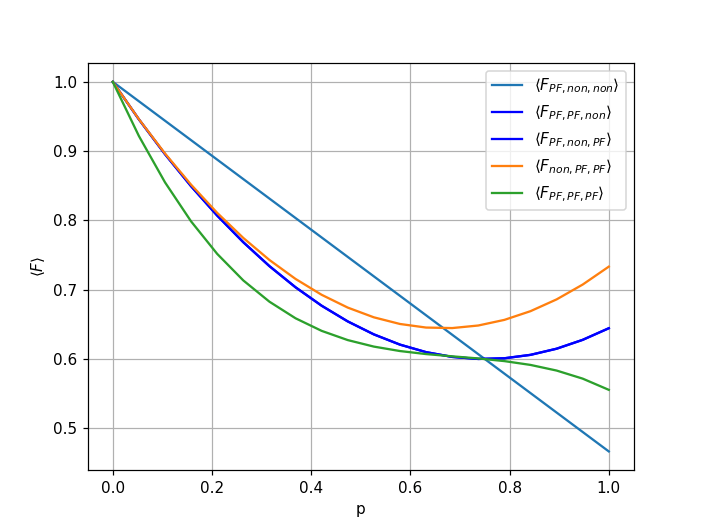} 
        \caption{Phase Flip Noise} 
        \label{fig 16(b)} 
        \vspace{2ex}
    \end{subfigure} 
    \begin{subfigure}[b]{0.5\linewidth}
    \centering
        \includegraphics[width=0.5\linewidth]{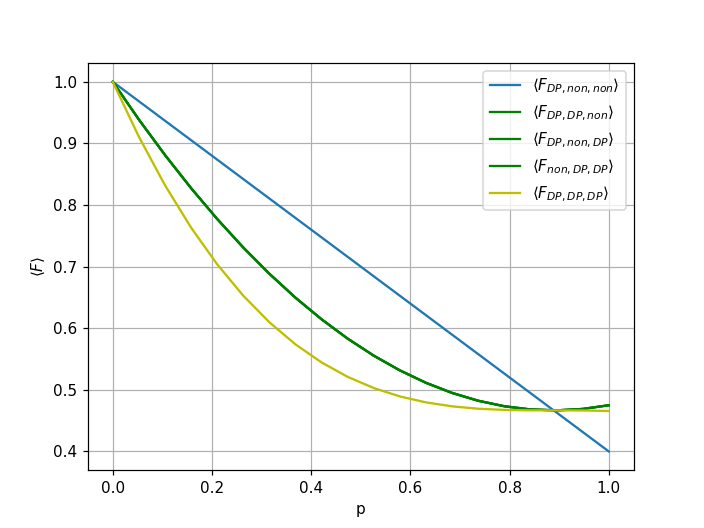} 
        \caption{Depolarization Noise} 
        \label{fig 16(c)} 
    \end{subfigure}
    \begin{subfigure}[b]{0.5\linewidth}
    \centering
        \includegraphics[width=0.5\linewidth]{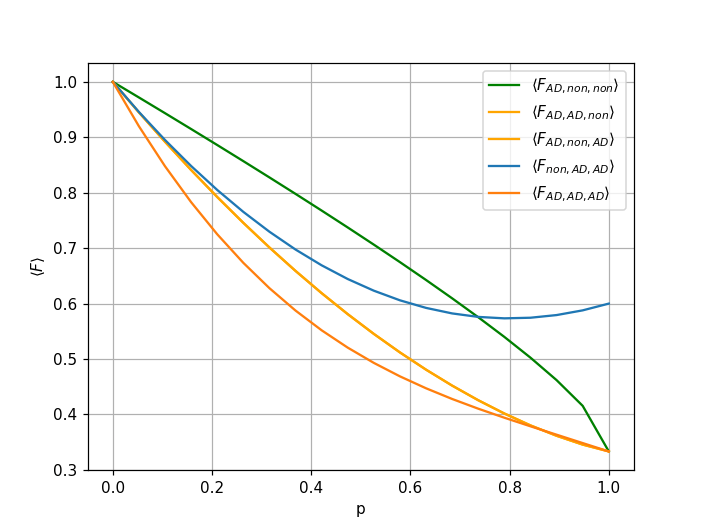}
        \caption{Amplitude Damping Noise} 
        \label{fig 16(d)} 
  \end{subfigure} 
\caption{Fidelity for each noise and it's combinations.}
\label{fig16} 
\end{figure}

FIG.(\ref{fig16}) illustration of average fidelity curve for different combination of all four noise, in which it shows fidelity when any noise say 'X' out of four noise acting input state only, acting both Alice state but not bob's channel state, acting Alice input state and Bob's channel state, acting both channel state respectively. The conclusion form FIG.(\ref{fig 16(a)} to \ref{fig 16(d)}) given in table(\ref{table:1}).

\begin{table}[ht!]
\centering
\caption{BF: Bit Flip, PF: Phase Flip, DP: Depolarization, AD: Amplitude Damping}
\begin{tabular}[t]{l c c}
\hline
FIG 16 & X & Fidelity\\
\hline
(a) &BF& $\langle F_{BF,non,non} \rangle < 
\langle F_{BF,BF,BF} \rangle < \langle F_{BF,BF,non} \rangle = \langle F_{BF,non,BF} \rangle = \langle F_{non,BF,BF} \rangle$\\
(b) &PF& $\langle F_{PF,non,non} \rangle < 
\langle F_{PF,PF,PF} \rangle < \langle F_{PF,PF,non} \rangle = \langle F_{PF,non,PF} \rangle < \langle F_{non,PF,PF} \rangle$\\
(c) &DP& $\langle F_{DP,non,non} \rangle < 
\langle F_{DP,DP,DP} \rangle < \langle F_{DP,DP,non} \rangle = \langle F_{DP,non,DP} \rangle = \langle F_{non,DP,DP} \rangle$\\
(d) &AD& $\langle F_{AD,non,non} \rangle = 
\langle F_{AD,AD,AD} \rangle = \langle F_{AD,non,AD} \rangle = \langle F_{AD,AD,non} \rangle < \langle F_{non,AD,AD} \rangle$\\
\hline
\end{tabular}
\label{table:1}
\end{table}

\section{Correlated Amplitude Damping(CAD)}\label{sec:Correlated Amplitude Damping(CAD).}

Uncorrelated amplitude damping refers to the noise acting on each qutrit independently whereas in correlated amplitude damping the noise acts simultaneously on the qutrits. When CAD acts on a state it changes into a different state with a lower amplitude.
Consider the entangled channel shared between Alice and Bob and we have to teleport the  information state  in Eq.(\ref{eq:qutritinput}) $|\psi_{in}\rangle = a|0\rangle + b|1\rangle + c|2\rangle$ from Alice to Bob. In the case of correlated amplitude damping, the channel state initially suffer amplitude damping noise mentioned in part 4 in a noisy channel and also  operators $A_{k}$   $(k\in {0,1,2})$ act on the channel that describes the correlation between two end of the channel.

The dynamics of the entangled state subjected to CAD noise can be expressed as the  quantum superposition $\epsilon_{CAD}$ acting on the initial state 

\begin{equation}
    \epsilon_{CAD} = (1-\eta)\sum_{i,j=0}^{2} {K_{ij} \rho K_{ij}^\dagger} + \eta\sum_{k=0}^{2}{A_{k} \rho A_{k}^\dagger}
\end{equation}
where $\eta$ is the correlated parameter and $0\le\eta \le 1$.\cite{li2019enhance}. When $\eta = 0$ we get uncorrelated amplitude damping channel and when $\eta = 1$ we get fully correlated amplitude damping channel. The Kraus operators corresponding to amplitude damping are given by $K_0 , K_1 ,K_2$ as given in part 4. $E_{ij}$ refers to the tensor product of the above Kraus operators. $A_{k}$ can be obtained by solving the Lindblad Master equation\cite{xu2022enhancing}. The terms $p_1, p_2$ in eq(85) gives the correlation strength.

\begin{equation}
 A_0 =\begin{bmatrix}
 I_{4}   &  &  & \\ 
   & \sqrt{1-p_2} &  & \\ 
   &  &  \ I_{3} & \\ 
   &  &   & \sqrt{1-p_1}
 \end{bmatrix}_{9\times 9}, A_1 = \begin{bmatrix} 
    \sqrt{p_1} & 0 & \dots & 0\\
    0 & 0 & \dots & 0 \\
    \vdots & \vdots & \ddots &  0\\
    0 &   0 & \dots    & 0 
\end{bmatrix}_{9\times 9},
 A_2 = \begin{bmatrix} 
    0 & \dots & 0 & \sqrt{p_1}\\
    0 & 0 & \dots & 0 \\
    \vdots & \vdots & \ddots & \vdots  \\\
    0 &   0 & \dots    & 0 
 \end{bmatrix}_{9\times 9}
\end{equation}

\begin{figure}
     \centering
     \begin{subfigure}[b]{0.5\textwidth}
         \centering
         \includegraphics[width=\textwidth]{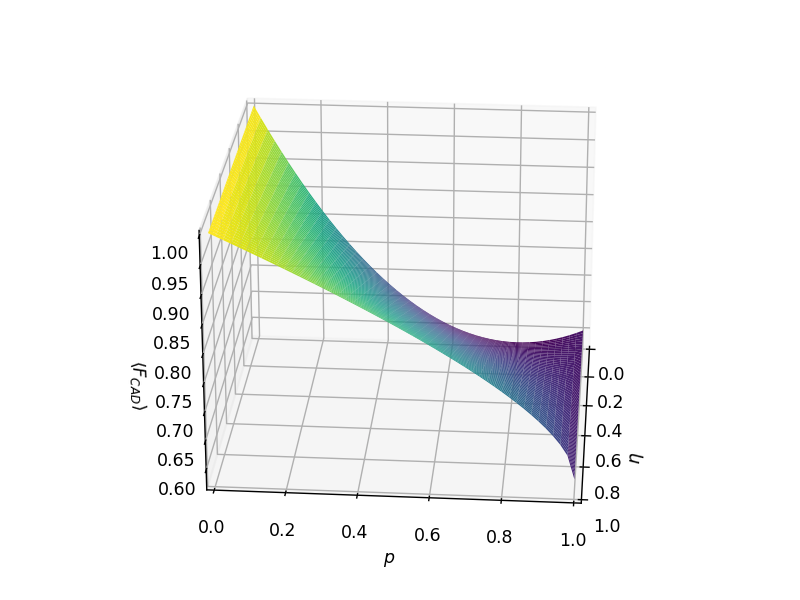}
         \caption{}
         \label{fig:17(a)}
     \end{subfigure}
     \begin{subfigure}[b]{0.41\textwidth}
         \centering
         \includegraphics[width=\textwidth]{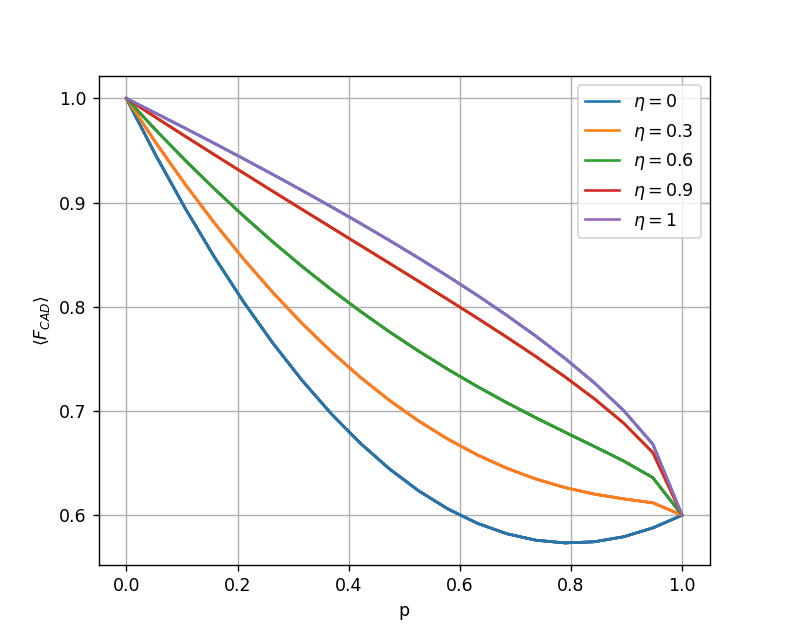}
         \caption{}
         \label{fig:17(b)}
     \end{subfigure}
\caption{Average fidelity of correlated amplitude damping}
\label{fig:17}
\end{figure}

When the standard procedure of teleportation is done, Bob gets the density matrix of the output state ($\rho_{out}$). The average fidelity can be obtained by the eq (23) given in part 3.
In terms of $\eta , p_A ,p_B, p_1, p_2$ the average fidelity can be written. The average fidelity as a function of damping probability  $p_A = p_B = p_1 = p_2 = p$ and correlation parameter $\eta$ is as shown in FIG.(\ref{fig:17(a)}) and its expression can be written as eq.(\ref{eq:86})
\begin{equation}
    F_{av}= -\frac{2\eta p^2}{3} + \frac{16\eta p}{15} + \frac{4\eta\sqrt{1-p}}{15} + \frac{2\eta(1-p)}{15} - \frac{2\eta}{5} + \frac{2p^2}{3} - \frac{16p}{15} + 1 
    \label{eq:86}
\end{equation}
FIG(\ref{fig:17(a)}) shows plot of average fidelity of correlated amplitude damping in channel state as function of noise probability $p$ and correlation parameter $\eta$ and FIG.(\ref{fig:17(b)}) is the plot of average fidelity of correlated amplitude damping in channel state as function of noise strength $p$ at different values of $\eta$. It shows that fidelity correlation amplitude damping is independent of $\eta$ as well as equal to fidelity of when amplitude damping acting in channel state described by the eq(\ref{eq:67}). 
\section{Conclusion}
We analyzed the effect of noise on Qutrit teleportation. Four different types of noises:Bit flip, phase flip, depolarization and amplitude damping acting on input state alone, channel states and on  the possible combination of these states are studied and fidelity in each case calculated. In the first scenario,when noise acts on the input state alone,it was observed that the fidelity is least. So it is better to keep the input state devoid of any kind of noise. If noise is unavoidable in input state, it is better to give the same noise in  Alice channel state or Bob's channel state for improved teleportation. In teleportation protocol if noise is certain in Alice end then it is better to keep Bob's end free from noise.Contrary to the above case, if we have the control over noise in Bob's end,it is finer to allow depolarization for Bit flip and depolarization noise in both Alice states, Bit flip for Phase flip respectively. We showed that if noise is unavoidable,then it is better to have them only on the channel state for increase in fidelity.This is because of the correlation effect between channel state. We also confirmed this behaviour for the amplitude damping noise by considering correlated amplitude damping in channel state.   In the last section we analysed all the possible combinations of each noise and we showed that if one have control of noise over any quantum state it is best to allow phase flip noise only on them for better teleportation. 


\bibliographystyle{unsrt}
\bibliography{ref}
\end{document}